\documentclass[pra,superscriptaddress,showpacs,preprint,aps,longbibliography]{revtex4-1}
\usepackage{epsfig}             
\usepackage{epstopdf}           
\usepackage{hyperref}           
\usepackage{color}
\usepackage{colortbl}
\usepackage{graphicx}
\usepackage{amssymb,amsmath}
\usepackage{subfigure}
\usepackage{caption}
\usepackage{enumerate}
\usepackage{gensymb}
\usepackage{url}

\begin{document}

\title[HHG]{Determination of the spectral variation origin in high-order harmonic generation in noble gases}
\author{V. E. Nefedova}
\affiliation{Institute of Physics of the ASCR, ELI-Beamlines project, Na Slovance 2, 182 21 Prague, Czech Republic}
\affiliation{Faculty of Nuclear Sciences and Physical Engineering CTU, Brehova 7, 115 19 Prague 1, Czech
Republic}

\author{M. F. Ciappina}
\email[]{marcelo.ciappina@eli-beams.eu}
\affiliation{Institute of Physics of the ASCR, ELI-Beamlines project, Na Slovance 2, 182 21 Prague, Czech Republic}

\author{O. Finke}
\affiliation{Institute of Physics of the ASCR, ELI-Beamlines project, Na Slovance 2, 182 21 Prague, Czech Republic}
\affiliation{Faculty of Nuclear Sciences and Physical Engineering CTU, Brehova 7, 115 19 Prague 1, Czech
Republic}

\author{M. Albrecht}
\affiliation{Institute of Physics of the ASCR, ELI-Beamlines project, Na Slovance 2, 182 21 Prague, Czech Republic}
\affiliation{Faculty of Nuclear Sciences and Physical Engineering CTU, Brehova 7, 115 19 Prague 1, Czech
Republic}

\author{J. V\'abek}
\affiliation{Institute of Physics of the ASCR, ELI-Beamlines project, Na Slovance 2, 182 21 Prague, Czech Republic}
\affiliation{Faculty of Nuclear Sciences and Physical Engineering CTU, Brehova 7, 115 19 Prague 1, Czech
Republic}

\author{M. Kozlov\'a}
\affiliation{Institute of Physics of the ASCR, ELI-Beamlines project, Na Slovance 2, 182 21 Prague, Czech Republic}
\affiliation{Institute of Plasma Physics ASCR, Za Slovankou 3, 182 00 Prague 8, Czech Republic}

\author{N. Su\'arez}
\affiliation{ICFO - Institut de Ciencies Fotoniques, The Barcelona Institute of Science and Technology, 08860 Castelldefels (Barcelona), Spain}

\author{E. Pisanty}
\affiliation{ICFO - Institut de Ciencies Fotoniques, The Barcelona Institute of Science and Technology, 08860 Castelldefels (Barcelona), Spain}

\author{M. Lewenstein}
\affiliation{ICFO - Institut de Ciencies Fotoniques, The Barcelona Institute of Science and Technology, 08860 Castelldefels (Barcelona), Spain}
\affiliation{ICREA, Pg. Llu\'{\i}s Companys 23, 08010 Barcelona, Spain}

\author{J. Nejdl}
\email[]{jaroslav.nejdl@eli-beams.eu}
\affiliation{Institute of Physics of the ASCR, ELI-Beamlines project, Na Slovance 2, 182 21 Prague, Czech Republic}
\affiliation{Institute of Plasma Physics ASCR, Za Slovankou 3, 182 00 Prague 8, Czech Republic}

\date{\today}
\pacs{32.80.Rm,33.20.Xx,42.50.Hz}

\begin{abstract} 
One key parameter in the high-order harmonic generation (HHG) phenomenon is the exact frequency of the generated harmonic field. 
Its deviation from perfect harmonics of the laser frequency can be explained considering (i) the single-atom laser-matter interaction and (ii) the spectral changes of the driving laser. In this work, we perform an experimental and theoretical study of the causes that generate spectral changes in the HHG radiation.
We measured the driving laser spectral shift after high harmonic generation in long medium using a correction factor to take into account the multiple possible HHG initiation distances along the laser path. We separate out the contribution of laser spectral shift from the resultant high harmonic spectral shift in order to elucidate the microscopic effect of spectral shift in HHG.
Therefore, in some cases we are able to identify the dominant electron trajectory from the experimental data. Our investigations lead to valuable conclusions about atomic dipole phase contribution to a high harmonic spectral shift. We demonstrate that the significant contribution of a long electron path leads to a high harmonic shift, which differs from that expected from the driving laser. Moreover, we assess the origin of the high-order harmonics spectral broadening and provide an explanation for the narrowest high harmonic spectral width in our experiment.
\end{abstract}

\maketitle

\section{Introduction}
 Development of coherent light sources in the extreme ultraviolet (XUV) spectral domain is of very high importance, because it allows researchers to carry out wavelength-limited imaging, observation and potentially control of various physical, biological and chemical phenomena at their natural spatial, nanometric, and temporal, sub-femtosecond, scales. The highly nonlinear interaction of intense femtosecond laser pulses with matter leads to the generation of harmonics of the driving laser light up to very high orders. This phenomenon, so-called high-order harmonic generation (HHG), became the flagship source of short wavelength coherent radiation~\cite{ferencreview}. A major benefit of sources based on HHG lies in their compactness and low cost, enabling wide availability for researchers. Besides its high degree of coherence, HHG radiation exhibits collimated beams~\cite{ditmire} with a Gaussian-like transverse energy distribution and a nearly diffraction-limited wavefront~\cite{gautier}, allowing for efficient focusing of the light onto samples. Furthermore, these properties facilitate the analysis of the source features and allow engineering of their temporal and spatial characteristics.
 Likewise, the polarization state of the emitted radiation can be controlled to a high degree~\cite{fleischer} (for a recent review see e.g.~\cite{mishaOptLet2017}).

 
During the last few decades significant efforts were put into the understanding of the main mechanisms driving the HHG process.  From a spectral point of view, HHG has been demonstrated over a very wide range of wavelengths, spanning from tens of nanometers to a few angstroms~\cite{taka}. Additionally, the intrinsic nature of the HHG phenomenon allows the generation of ultrashort pulses; indeed, it was shown that these sources can reach the attosecond (as) temporal range (1 as = 10$^{-18}$ s)~\cite{paul, chang}, allowing an insight into the ultrafast electron dynamics inside matter \cite{Corkum_Nature}.
Other possible applications of these ultrashort coherent XUV sources include following transient elementary processes~\cite{tunnel}, scrutinizing ultrafast magnetization effects~\cite{vodun} and monitoring molecular processes in real time~\cite{zhou}.
 Depending on the technology of the laser driver, HHG sources can be operated in a single-shot mode or at very high repetition rates.  The former makes it possible to conduct pump-probe experiments with samples that are non-renewable~\cite{tzallas,twophoton}, while the later favors studies (such as photoelectron spectroscopy) where accumulation of a low intensity signal is required.

It is therefore important to find tools able to control the frequency of the XUV ultrashort pulses generated through HHG. The current contribution provides a detailed study of the HHG frequency variation, which could be advantageously used for time-resolved ultrafast experiments. At a single-atom level, the HHG process can be easily explained by a semiclassical description, namely invoking what is known as the ``three-step model": first, an atom is ionized by the strong electric field of the laser, so the electron wave packet can leave the atomic potential via tunneling; next, the free electron is accelerated by the laser field, defining the second stage of the sequence, and finally, when the laser electric field reverses sign, the electron can recombine with its parent ion, emitting the energy obtained during its journey in the laser-dressed continuum in the form of XUV radiation~\cite{corkum,kulandernato,kulander1992}. 

From a macroscopic viewpoint, the individual fields emitted by atoms located at different positions in the propagation direction need to be in phase, so that their respective electric fields can add up constructively, allowing the growth of the XUV intensity. The wavevector mismatch, defined as $\Delta k= q k_1- k_q$, where $q$ is the harmonic order and $k_1$ and $k_q$ are the wavevectors of the fundamental driving field and the one of the $q^{\mathrm{th}}$ harmonic order, respectively, consists of contributions from plasma $\Delta k_{p}$ and neutral atoms $\Delta k_{n}$. There is also a mismatch due to the Gouy phase shift $\Delta k_{G}$ and the atomic dipole $\Delta k_{d}$, which originates from the intensity gradient of the incident beam and depends on the electron trajectory (short or long) responsible for the emission. 
One of the areas of intense research encompasses the development and implementation of configurations with
larger interaction lengths - the length over which the two fields are phase-matched, known as the coherence length and defined as $L_{\mathrm{coh}}=\frac{\pi}{\Delta k}$. The condition, which minimizes the wave vector mismatch and turns $\Delta k$ close to zero, is called phase-matching~\cite{Ruchon,Delfin}.  

Considerable attempts have been made to obtain control of the different quantum path contributions with a low driving laser intensity~\cite{Merdji}. In this work, we use a method that allows us to determine the origin of the spectral variation of individual harmonics when the driving laser intensity increases. We perform a real-time monitoring of the driving laser's spectral properties, which provides a beneficial approach for its control during the HHG process. The post-interaction IR spectra are recorded on-axis, simultaneously with the HHG spectrum. The obtained data are then used in order to discriminate the contributions due to the driving laser wavelength shift, from the intrinsic ones that are inherent to every HHG phenomenon.

This paper is organized as follows: in the next section, Sec.~II, we describe the experimental setup. Section III is devoted to the theoretical framework, including all the sources for the phase variation of HHG and Sec.~IV is focused on the results and discussion. Here we include the determination of the intensity-dependent quantum-path phase coefficients and a comparison between our theoretical predictions with the experimental data. Finally, in Sec.~V, we summarize the main ideas, present our conclusions.

\section{Experimental setup}
The HHG experiment was carried out using a 10 Hz Ti:Sapphire laser system, wavelength centered at about $\lambda_{\mathrm{IR}}=810$ nm with a pulse energy from 50 mJ to 125 mJ. Additionally, the pulse duration was estimated to be 50 fs full-width-half-maximum (FWHM) using an autocorrelation technique. The laser beam was loosely focused by a spherical mirror with a focal length of $f=5$ m into a gas cell filled with a noble gas (argon, neon or helium), as shown in Fig.~1.
\begin{figure}[ht]
\begin{center}
\includegraphics[width=0.85\textwidth]{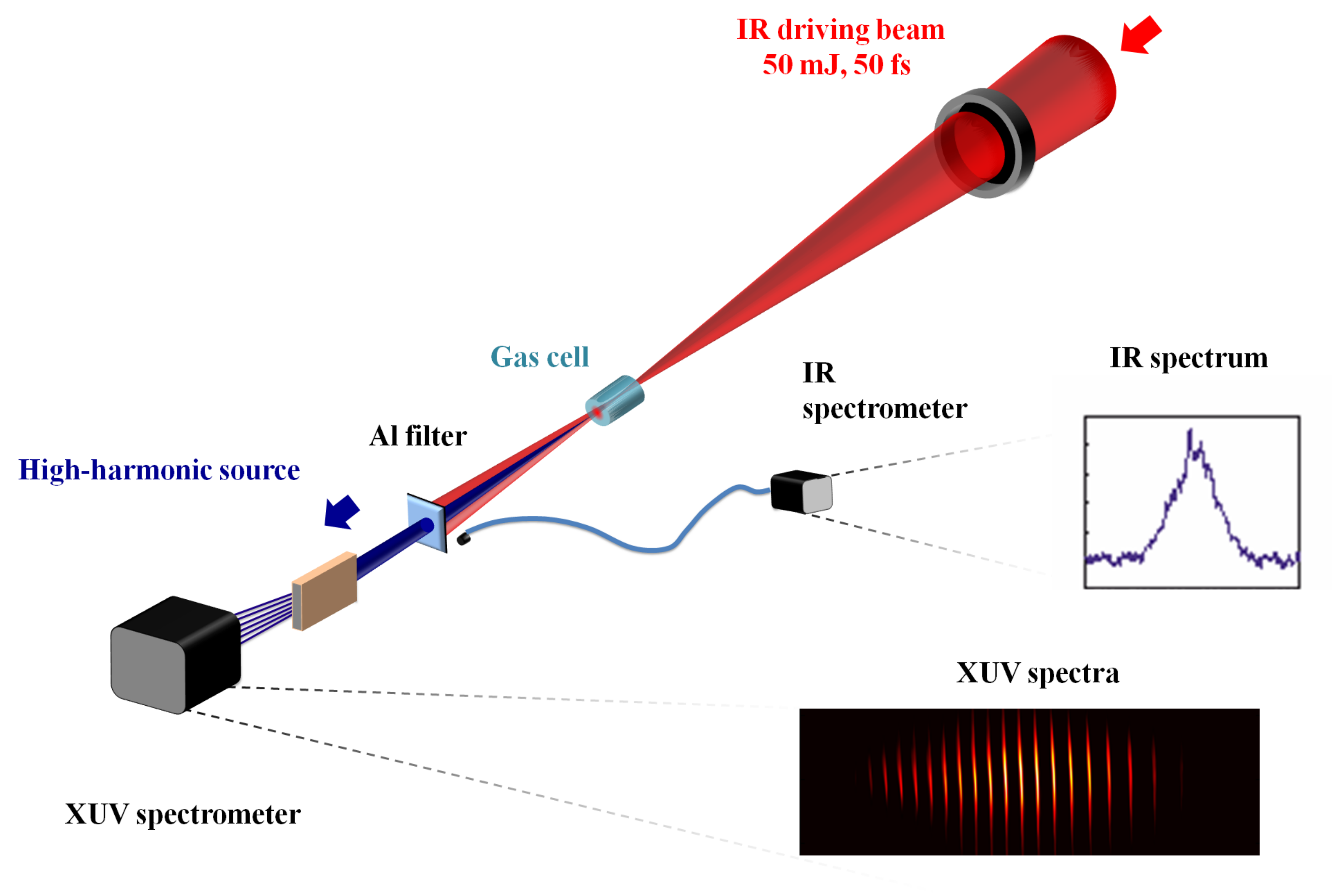}
\caption{(color online) Scheme of the experimental setup. By focusing an IR driving laser into a gas cell filled with rare gas, a harmonic beam is generated. An IR spectrometer records the driving laser spectrum after the interaction with the gas and a XUV spectrometer detects the produced high-order harmonic spectra.}
\label{Figure 1}
\end{center}
\end{figure}

The incident laser beam diameter ($D$) (and thus the transferred energy $E$) is varied by changing an iris aperture diameter providing a range of $F_{\#}$ ($F_{\#}=f/D$) from 150 to 400. The fundamental IR beam was focused into the gas cell to produce high-order harmonics, which follow the laser propagation direction. A thin metallic transmission filter was placed about 5 m away from the HHG source at the entrance of the diagnostics chamber in order to reject the IR radiation. A flat-field spectrometer, with a reflective concave diffraction grating and a back-illuminated CCD camera, was used for recording the HHG spectra. A fiber spectrometer was employed for obtaining the fundamental beam's spectrum after the interaction with the gas. The fiber connected to the spectrometer collected the scattered IR light from the metallic filter
(16 mm in diameter). This procedure allows us to select the central part of laser beam and to perform on-axis measurements.
The HHG yield optimization was done by adjusting the iris diameter, thus varying the laser energy on the target, the $F_{\#}$ and consequently the IR laser pulse intensity at focus. Since the intensity gradient of the field is small in our loose focusing geometry, we neglect contributions from Gouy phase shift $\Delta k_{G}$ and atomic dipole phase $\Delta k_{d}$ when we change the iris diameter. Besides, when we increase the laser driving intensity, the influence of the medium ionization on the phase-matching terms $\Delta k_{p}$ and $\Delta k_{n}$, becomes dominant. 
The optimal conditions were found to be $F_{\#}=300$, $E=50$ mJ for HHG in argon, the gas cell length and pressure were $L=10$  cm and $p=5$ mbar ($N_{a}$ =1.25 $\times$ 10$^{23}$ atoms/m$^3$), respectively. Here, the beam focus was at the center of the cell. On the other hand, for HHG in neon we have $F_{\#}=240$, $E=85$ mJ, $L=4.4$ cm and $p=20$ mbar ($N_{a}$ =5 $\times$ 10$^{23}$ atoms/m$^3$). For this case the end of the gas cell was about 1.2 cm after the beam focus. Finally, $L=3.6$ cm and the end of the gas cell was placed at the beam focus for HHG in helium. Here the optimal conditions were $F_{\#}=200$, $E=125$ mJ and $p=85$ mbar ($N_{a}$ =2.13 $\times$ 10$^{24}$ atoms/m$^3$).

\subsection{Driving laser intensity estimation}
The driving laser intensity for different iris aperture sizes is estimated from the measured laser energy, pulse duration and the focal spot intensity distribution. The intensity distribution of the attenuated beam is recorded directly by a CCD camera placed at the laser beam focus. 
On the other hand, the determination of the laser intensity $I_{\mathrm{cutoff}}$, one needs for generation of a high-order harmonic of energy $\hbar\omega_{\mathrm{cutoff}}$, is done by applying the semiclassical formula~\cite{maciej1} 
\begin{equation}
\label{cutoff}
{\hbar\omega_{\mathrm{cutoff}}=I_p+3.17U_p},
\end{equation}
where $I_p$ is the ionization potential of the atom under consideration and $U_p$ the ponderomotive energy, which can be expressed as $U_p$ [eV] $=9.33\times10^{-14}I_{\mathrm{cutoff}}$ [W/cm$^2$] $\lambda_{\mathrm{IR}}^2$ [$\mu$m$^2$], $\lambda_{\mathrm{IR}}$ being the laser central wavelength. 
Finally, a particular estimated peak intensity is divided by a constant coefficient in order to get values closer to $I_{\mathrm{cutoff}}$.  The difference between the vacuum peak laser intensity and the $I_{\mathrm{cutoff}}$  mainly originates from the non gaussian temporal shape of the driving laser pulse as well as aberrations of the focused beam. All the laser intensity values used in our calculations are computed keeping that coefficient constant.

\section{Theory}
The harmonic phase originates from two components: one coming from the driving laser field phase, defined as $\phi_{\mathrm{IR},q}=q \phi_{\mathrm{IR}}$ \cite{Wahlstrom, Rae93}, where $\phi_{\mathrm{IR}}$ is the driving laser phase and $\phi_{\mathrm{IR},q}$ is the phase of the $q^{\mathrm{th}}$ harmonic, and another associated with the atomic dipole phase $\phi_{\mathrm{dipole}}$. Since the harmonics are generated by a laser pulse, the intensity of which varies in a temporal domain as $I(t)$, the dipole phase leads to frequency shift and chirp of the harmonic pulse. The resulting total phase of the generated harmonics $\phi_q(t)$ can then be written as:
\begin{equation}
\label{total_phase}
\phi_q(t) = \phi_{\mathrm{IR},q}(t)+\phi_{\mathrm{dipole}}(t).
\end{equation}

The instantaneous frequency of the $q^{\mathrm{th}}$ harmonic is  
\begin{equation}
\label{total_frequency}
\omega_{\mathrm{inst, q}}(t)=\omega_q+\frac{\partial \phi_{q}(t)}{\partial t}=\omega_q+\Delta\omega_q(t)=\omega_q+\Delta\omega_{\mathrm{IR},q}(t)+\Delta\omega_{\mathrm{dip}}(t),
\end{equation}
where $\Delta\omega_{\mathrm{IR},q}(t)$ and $\Delta\omega_{\mathrm{dip}}(t)$ are frequency variations due to driving laser frequency variation and atomic dipole contribution, respectively.
The central wavelength shift can be expressed as
$\Delta\lambda_{q}=\frac{\lambda_{q}^2}{2\pi c}\Delta\omega_q$.
The contribution from the IR field, $\phi_{\mathrm{IR},q}$, will be treated in Sec.~III.A, while the contribution from the intrinsic dipole phase, $\phi_{\mathrm{dipole}}$, will be studied in Sec.~III.B. 

\subsection{Driving laser frequency variation}
One can estimate the $q^{\mathrm{th}}$ harmonic frequency, originating from the laser driving field $\phi_{\mathrm{IR},q}$ [see Eq.~(\ref{total_phase})], by considering plasma effects on the driving laser pulse. The spectral shift $\Delta \omega_{\mathrm{IR}}$ in the IR field due to plasma is transferred to a spectral shift $\Delta \omega_{p,q}$ of the $q^{\mathrm{th}}$ harmonic  via $\Delta \omega_{p,q}= q\times \Delta \omega_{\mathrm{IR}}$ \cite{Wahlstrom, Rae93}.
We will compute the expected shift for the qth harmonic in two steps: first, only the shift of the driving laser is considered; second, the shift transferred to the harmonic field is estimated taking into account the absorption in the medium. The laser pulse phase determined by modulation due to temporal variation of refractive index $n(t)$ after propagation through a medium of length $L$ is
\begin{equation}
\label{}
\phi_{\mathrm{IR}}(t)=-\frac{2\pi}{\lambda_{\mathrm{IR}}}\int_0^L n(t) dz
\end{equation}
leading to plasma-induced frequency shift as
\begin{equation}
\label{spectral_shift_IR}
\Delta \omega_{\mathrm{IR}}(t)=\frac{\partial \phi_{\mathrm{IR}}(t)}{\partial t}=-\frac{2\pi}{\lambda_{\mathrm{IR}}}\int_0^L \frac{\partial n(t)}{\partial t} dz,
\end{equation}
where, $n=\sqrt{1+\chi_p+\chi_n}$ devotes the refractive index due to plasma and neutral atoms, $c$ is the speed of light. The electric susceptibility due to plasma is $\chi_{p}=-\frac{\omega_p^2}{\omega_0^2}$, where $\omega_p$ is the plasma frequency $\omega_p=\sqrt{\frac{N_e e^2}{\epsilon_0 m_e}}$, with $e$ the elementary electric charge, $\epsilon_0$ the vacuum permittivity, $m_e$ the electron mass and $\omega_0$ the central angular frequency of the driving laser. The free electron density is $N_e=\frac{\eta N_{\mathrm{atm}}p}{p_0}$ with $\eta$ the ionization probability, $N_{\mathrm{atm}}$ the particle density of an ideal gas at standard temperature and pressure (STP),  $p_0 =1$ atm and $p$ the actual gas cell pressure. Moreover, for neutral atoms one obtains $\chi_{n}=\frac{N_a e^2}{\epsilon_0m_e(\omega_r^2-\omega_0^2)}$ \cite{chang2011}, where $\omega_r$ is the closest resonant frequency and the neutral atom density is given by $N_a=\frac{N_{\mathrm{atm}}p(1-\eta)}{p_0}$. Thus, both plasma and neutral atom contributions to the refractive index depend on the initial atomic density as well as on $\eta$. 

It can be inferred from Eq.~(\ref{spectral_shift_IR}), that the spectral shift is proportional to (i) the medium length $L$, (ii) the ionization rate $\frac{\partial \eta}{\partial t}$ and (iii) the particle density $N_a$, through the refractive index $n$. When the medium is strongly ionized, the plasma effect introduces a blueshift in the IR spectrum~\cite{leblanc}. 
\par The main advantage of our study is the possibility to distinguish between the effect of the fundamental laser spectral shift and the effect coming from the intrinsic atomic dipole in the resulting high harmonic spectral shift. This is possible, since we can determine IR contribution by directly measuring its spectrum after propagation through the gas cell.

\par We measure the driving laser spectra after the pulse propagates through the entire gas medium of length $L$. However, as high-order harmonics are generated from numerous positions throughout medium, it is important to estimate the effective contribution from the detected IR spectral shift to a resultant high harmonic spectral shift. In order to take into account this fact, we perform a correction for calculation of $\Delta \lambda_{\mathrm{IR}}$ in the resulting HHG spectral shift. For a simplified consideration, we assume a one-dimensional model and take into account that the IR laser acquires a spectral shift as $\Delta\lambda_{\mathrm{IR}}(z)=\frac{\Delta\lambda_{\mathrm{IR}}}{L}z$. For high order harmonics, the phase matching conditions are fulfilled and medium is considered as homogeneous, the high-order harmonic signal develops along a medium of length $L$ as ~\cite{Constant}:
 \begin{equation}
I_q \propto \left| \left[\int_{0}^{L_{}} E_q(z, \omega) \exp\left(-\frac{L_{}-z}{2L_{\mathrm{abs}}}\right) \exp (i\phi_q(z)) dz \right]\right|^2,
\label{I_q}
 \end{equation}
where $E_q(z, \omega)$ is the amplitude of the atomic response around the harmonic frequency $\omega_q$ and $\phi_q(z)$ is its phase, $z$ is the distance along the propagation direction and $L_{\mathrm{abs}}$ the absorption length, defined as $L_{\mathrm{abs}}=\frac{1}{\sigma N_a}$ ($\sigma$ is the photoabsortion cross-section).  We estimate $L_{\mathrm{abs}}$ =10 mm for argon, $L_{\mathrm{abs}}$ = 3.7 mm for neon and $L_{\mathrm{abs}}$ = 10.6 mm for helium at the pressures given in Section~II. 
Next, we need to evaluate the role of the emitters placed at different points in medium considering that the shift of the IR, $\Delta \lambda_{\mathrm{IR}}$, is acquired along the entire medium. We built up a simple model based on Eq.~(\ref{I_q}), assuming the harmonics are being generated uniformly along the medium. Moreover, we assume that an emitter placed in a position $z$ produces only monochromatic radiation corresponding to the actual laser frequency. Consequently, the spatial profile of the harmonic intensity, which is thus driven only by the exponential term, is directly imprinted in the spectrum. 
In order to be consistent with the experimental procedure for estimation of central wavelength taking a median value, for the next analysis we consider the correction factor $k$ given by the median of the profile as:
 \begin{equation}
k=1-\frac{2\, L_{\mathrm{abs}}}{L} \ln\left(\frac{2}{\exp\left(-\frac{L}{2\, L_{\mathrm{abs}}}\right)+1}\right),
\label{k_factor}
 \end{equation}
this value is already renormalised according to the transition from the spatial into the spectral domain. 
%
Finally, the effective high-order harmonic spectral shift due to the laser wavelength shift reads as $\Delta \lambda_{p,q}= k \Delta \lambda_{\mathrm{IR}}/q$. Using our experimental parameters we find that $k=0.86$ for argon, $k=0.88$ for neon and $k=0.72$ for helium.

\subsection{Atomic dipole phase}

The atomic dipole moment is the observation, that is used to quantify the radiation generated by
the $q^{\mathrm{th}}$ harmonic at the single-emitter level. Its phase is determined by the Volkov classical action $S_V(t,t')=I_p(t-t') + \frac12 \int_{t'}^t (\mathbf p+ \mathbf A(\tau))^2 \mathrm d \tau$ acquired by the laser-ionized free electron with drift momentum $\mathbf p$ at time $t'$ oscillating in the driving laser field until its recollision at time $t$, as $\phi_\mathrm{dipole} = S_V(t,t')-q\omega_0 t$, i.e.\ by the value of the action along the most relevant semiclassical trajectory~\cite{maciej1, Lewenstein1,HHGTutorial}, where  $\mathbf A(t)=-\int \mathbf E(t)\mathrm dt$ is the laser's vector potential.

This phase is therefore governed by the time integral of the kinetic energy of the electron during its journey, which increases with the mean velocity, and therefore with the laser intensity. 

The above explanation gives a straightforward physical picture of how the intrinsic dipole phase is acquired from a single emitter. In a simple approximation, the resulting phase shift is linear with the intensity,
\cite{Catoire}
\begin{equation}
\phi_\mathrm{dipole} = \phi_\mathrm{dipole}(I) \approx -\alpha I + \mathrm{const},
\end{equation}
where the quantum-path phase coefficient $\alpha = \frac{\partial\phi_\mathrm{dipole}}{\partial I}$ determines the core dependence of the harmonic phase on the laser intensity. It is characteristic for each generation regime, i.e. the long and short trajectories in plateau and the single trajectory in cutoff. The spectral shift can be also directly determined from a saddle-point calculation in SFA. These are discussed in detail in Sec.~IV.C.

\section{Results and discussion}

\subsection{Phase-matching}
 The major part of the XUV photon flux, during HHG, is produced provided that the phase-matching conditions are fulfilled. The medium refractive index $n$ is strongly linked to the concentration of both free electrons $N_e$ and neutral atoms $N_a$, that in turn are dictated by the ionization probability $\eta$. 
\begin{figure}[ht]
  \includegraphics[width=1\textwidth]{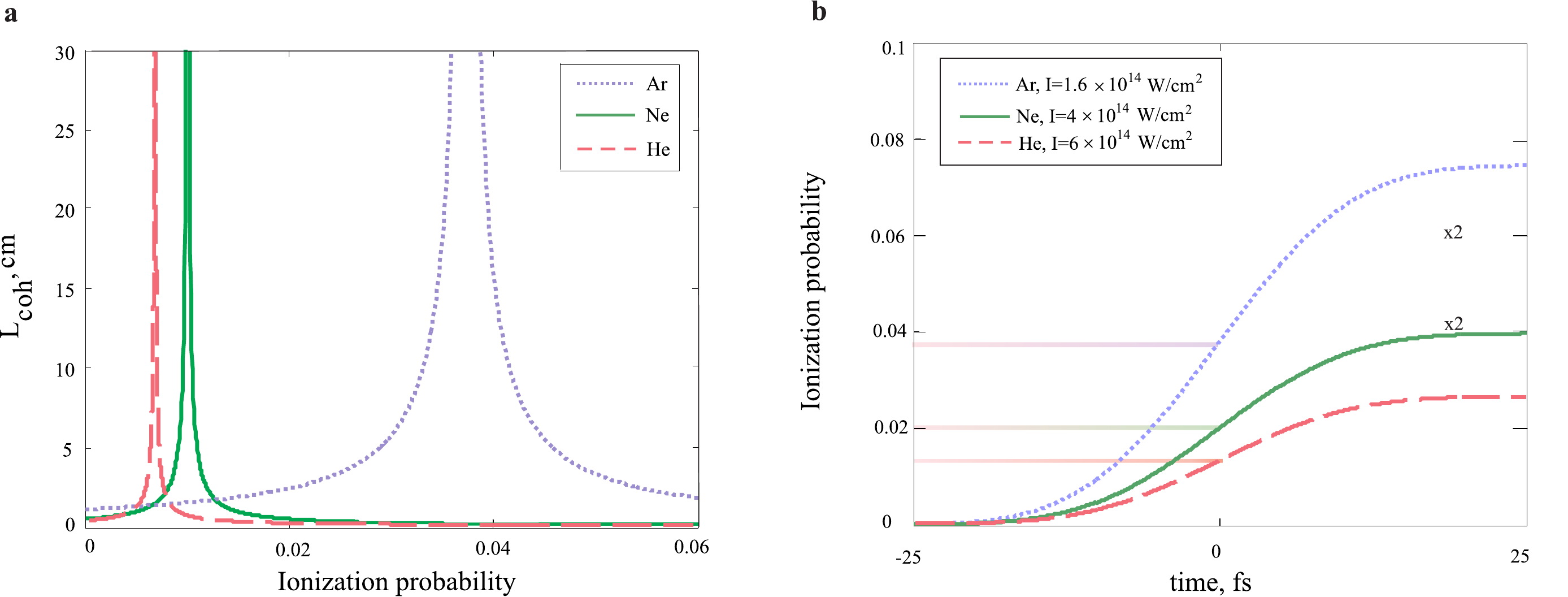}
    \caption{(a) Coherence length $L_{\mathrm{coh}}$ as a function of $\eta$ for the 25$^{\mathrm{th}}$ harmonic generated in argon (blue dotted line); the 57$^{\mathrm{th}}$ harmonic in neon (green solid line) and the 65$^{\mathrm{th}}$ harmonic in helium (red dashed line); The calculation was done for particle densities, which corresponds to our experimental conditions (Sec.~II); (b) ionization probabilities $\eta$ calculated for argon (blue dotted line), neon (green solid line) and helium (red dashed line). The horizontal lines define the phase-matched $\eta$ values. The laser peak intensity values $\hat{I}_{PM}$, where phase-matching occurs at the peak of our 50 fs pulse, are provided in the legend.
}\label{Figure2}
\end{figure}

Using $\Delta k_q=\frac{q \omega_0}{c} \Delta n_{\mathrm{pl,at}}$, where $\omega_0=\frac{2\pi c}{\lambda_{\mathrm{IR}}}$ and $\Delta n_{\mathrm{pl,at}}=\Delta n_{\mathrm{IR}}-\Delta n_{q}$, $\Delta n_{\mathrm{IR}}$ and $\Delta n_q$ being the refractive index variations due to plasma and neutral atoms for the driving laser and the $q^{\mathrm{th}}$ order harmonic, respectively. For estimation of phase-matching ionization level we neglect variation of the refractive index of high-order harmonics $\Delta n_{q}$, since it is significantly less than the variation of refractive index of the driving laser $\Delta n_{\mathrm{IR}}$. The resulting phase-matching ionization level $\eta_{PM}$ at which plasma and neutral atoms dispersion cancel each other~\cite{chang2011} can be estimated as:
\begin{equation}
\eta_{PM}\approx \left(1+\frac{{N}_{\mathrm{atm}} r_e \lambda_{\mathrm{IR}}^2}{2\pi\Delta n} \right)^{-1}
\end{equation}
where  $\Delta n$ is a change of noble gas refractive index at one atmospheric pressure, $N_{\mathrm{atm}}$ is the particle density of an ideal gas at STP conditions and $r_e$ the classical electron radius.

We plot the coherence length $L_{\mathrm{coh}}$ as a function of $\eta$ in Fig.~2(a), which was calculated for particle densities corresponding to our experimental conditions (Sec.~II). As can be seen, there is an optimal value $\eta_{PM}$, corresponding to an ideal phase-matching, of about $\sim 3.8$\% for the 25$^{\mathrm{th}}$ harmonic generated in argon with $\lambda_{\mathrm{IR}}=810$ nm, and about $\sim 1$\% for the 57$^{\mathrm{th}}$ harmonic in neon. Finally,  we find that $\eta_{PM} \sim 0.66$\% for the 65$^{\mathrm{th}}$ harmonic generated in helium. 
%
%
%

We can then define a laser intensity value $\hat{I}_{PM}$, which corresponds to the generation of phase-matched high-order harmonics at the peak of the laser pulse. The static ionization rates in the tunneling and barrier-suppression regimes are computed using the empirical formula from Ref.~\cite{tonglin}.  
Figure 2(b) depicts the time-dependence of $\eta$ for HHG in argon, neon and helium at different laser peak intensity values $\hat{I}_{PM}$ for a 50 fs long laser pulse with Gaussian distribution. The corresponding values of $\eta_{PM}$ are indicated by coloured horizontal lines.

\subsection{Determination of the intensity-dependent quantum-path phase coefficients}

To determine the values of the quantum-path phase coefficient $\alpha$, we calculate the harmonic dipole following the standard SFA formalism~\cite{maciej1, Lewenstein1,HHGTutorial}, in the form:
\begin{equation}
\mathbf D(q\omega_0) = \sum_s d(t_\mathrm{st},t'_\mathrm{st})e^{-iS_V(t_\mathrm{st},t'_\mathrm{st})+iq \omega_0 t_\mathrm{st}},
\end{equation}
where $t_\mathrm{st}$ and $t'_\mathrm{st}$ are the saddle-point solutions for recollision and ionization times at which the action $S_V(t,t')-q \omega_0 t$ is stationary, and we sum for a single ionization burst over the first six quantum trajectories~\cite{miloandbecker}. As can be seen, there are several return paths, which correspond to trajectories that spend increasingly long times oscillating in the continuum before recombining with the ion. Although some paths might be closed at a given intensity for a fixed harmonic order, as the returning electron does not yet have enough energy to produce those harmonics. As the excursion time spent in the continuum increases, so does the phase coefficient $\alpha$, so the first two returns (so-called short and long trajectories~\cite{HHGTutorial,miloandbecker}) have the lowest values of $\alpha$, with higher-order return producing a more sensitive phase dependence on the intensity.

\begin{figure}[htb]
  \includegraphics[width=1\textwidth]{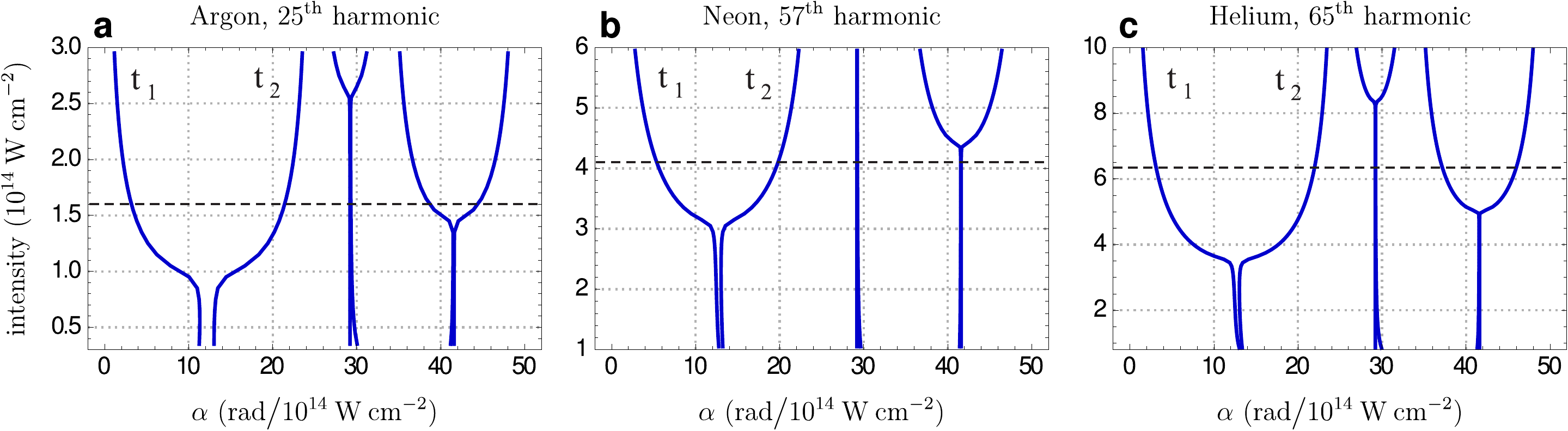}
    \caption{Plot of the quantum-path phase coefficient $\alpha$ computed using the SFA model for (a) the 25$^{\mathrm{th}}$ harmonic in argon, (b) the $57^{\mathrm{th}}$ harmonic in neon and (c) the 65$^{\mathrm{th}}$ harmonic in helium. The horizontal black dashed lines indicate the intensity level $\hat{I}_{PM}$, at which phase-matching radiation is produced from the laser pulse peak.
}
		  	\label{alpha_plot}
		\end{figure}

The phase of the emitted harmonics, then, is given by $\phi_\mathrm{dipole}=\Re(S_V(t_s,t'_s)-q \omega_0 t_s)$ and the $\alpha$ coefficient is obtained by numerical differentiation of this phase with respect to the intensity. The obtained $\alpha$ values as a function of laser intensity are shown in Fig.~3 for the 25$^{\mathrm{th}}$ harmonic in argon [Fig.~3(a)], the 57$^{\mathrm{th}}$ harmonic in neon [Fig.~3(b)] and the 65$^{\mathrm{th}}$ harmonic in helium [Fig.~3(c)].
 
 In our analysis we consider a piece-wise dependence of $\alpha$ with the laser intensity and only the first two trajectories are relevant (one short and one long). We identify them according to the their respective recombination times, $t_1$ and $t_2$. 
 
\subsection{Connection of the atomic dipole spectral shift with the intensity slope}
We can predict the behavior of the expected spectral shift due to the atomic dipole phase. In the next analysis, we consider $\alpha(I)$, that provides a more accurate description of the phase variation according to Section IV.B. 
The frequency variation due to the atomic dipole phase can be calculated by~\cite{Lewenstein1, Murakami2005}
\begin{equation}
\Delta\omega_{\mathrm{dip}}(t)=-\frac{\partial \, \phi_{\mathrm{dipole}}(t)}{\partial\, t}
=-\frac{\partial I}{\partial t}\left(\frac{\partial \alpha(I)}{\partial I}I +\alpha(I)\right).
\label{del_omega_dip}
\end{equation}

First, we calculate the expected shift originating from the atomic dipole phase following the theory of Section III.B and Eq.(~\ref{del_omega_dip}) with the $\alpha$ coefficients computed according to the procedure described in Section IV.B. 
We assume the high-order harmonic radiation is generated mostly in the temporal region where the laser intensity provides an ionization degree $\eta_{PM}$ corresponding to a phase-matching level (see Section IV.A); we take the amount of frequency variation $\Delta\omega_{\mathrm{dip}}$ due to the laser intensity slope at these points and calculate the expected wavelength shift as $\Delta\lambda_{\mathrm{dip}}=\frac{\lambda_q^2}{2\pi c}\Delta\omega_\mathrm{dip}$, where $\Delta\omega_\mathrm{dip}$ is given by Eq.~(\ref{del_omega_dip}).
 \par
\begin{figure}[htb]
  \includegraphics[width=1\textwidth]{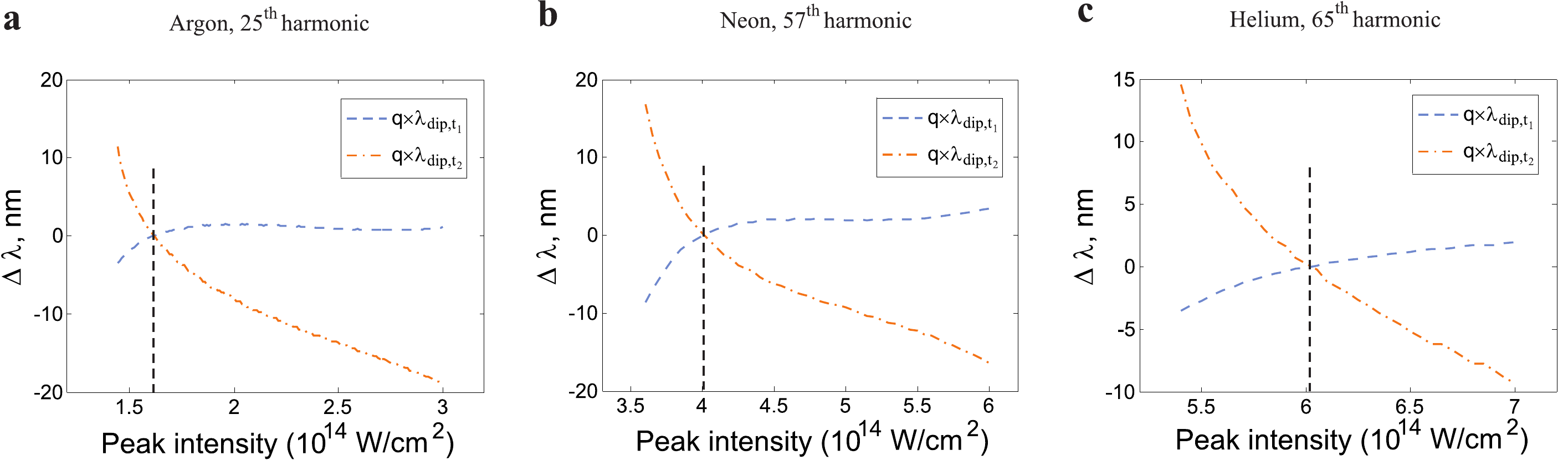}
    \caption{Spectral wavelength shift due to atomic dipole phase calculated according to Eq.~(\ref{del_omega_dip}) for HHG produced in phase-matching time for (a) argon, (b) neon and (c) helium. The $\alpha$ values are taken from Fig.~3. The dashed black lines correspond to the intensity level $\hat{I}_{PM}$, at which phase-matching radiation is produced from the laser pulse peak.
}
\label{Fig:shift_alpha}
\end{figure}

 Fig.~\ref{Fig:shift_alpha} demonstrates calculated high harmonic spectral shift due to the atomic dipole phase. In here, the short and long trajectories for the first return are considered. 
As can be inferred, 
%
%
 $\alpha$ variation with intensity in Eq.~(\ref{del_omega_dip}) for a short trajectory leads to a noticeable blueshift at the low laser intensities (i.e.~in conditions where $I_{peak}<\hat{I}_{PM}$) and minor and slowly varying redshift at high laser intensities ($I_{peak}>\hat{I}_{PM}$). In contrast, spectral shift due to the atomic dipole, taking into account a long electron trajectory predicts visible redshift at the low laser intensities ($I_{peak}<\hat{I}_{PM}$) and strong blueshift at high intensities ($I_{peak}>\hat{I}_{PM}$). These data will assist in the identification of the dominant electron trajectory during experiment, as can be seen in next Section.

\subsection{The measured and calculated spectral shifts}

We study the intensity-dependent central wavelength shift of both the driving laser and particular high-order harmonics. In order to identify the origin of the high-order harmonic spectral shift following Eq.~(\ref{total_frequency}), both the frequency shift due to atomic dipole as well as the spectral shift of the driving IR laser contribution, should be taken into account.
The predicted wavelength shifts due to the atomic dipole phase are shown in Fig.~\ref{Fig:shift_alpha} for argon, neon and helium atoms.  

\begin{figure}[ht]
\begin{center}
\includegraphics[width=1\textwidth]{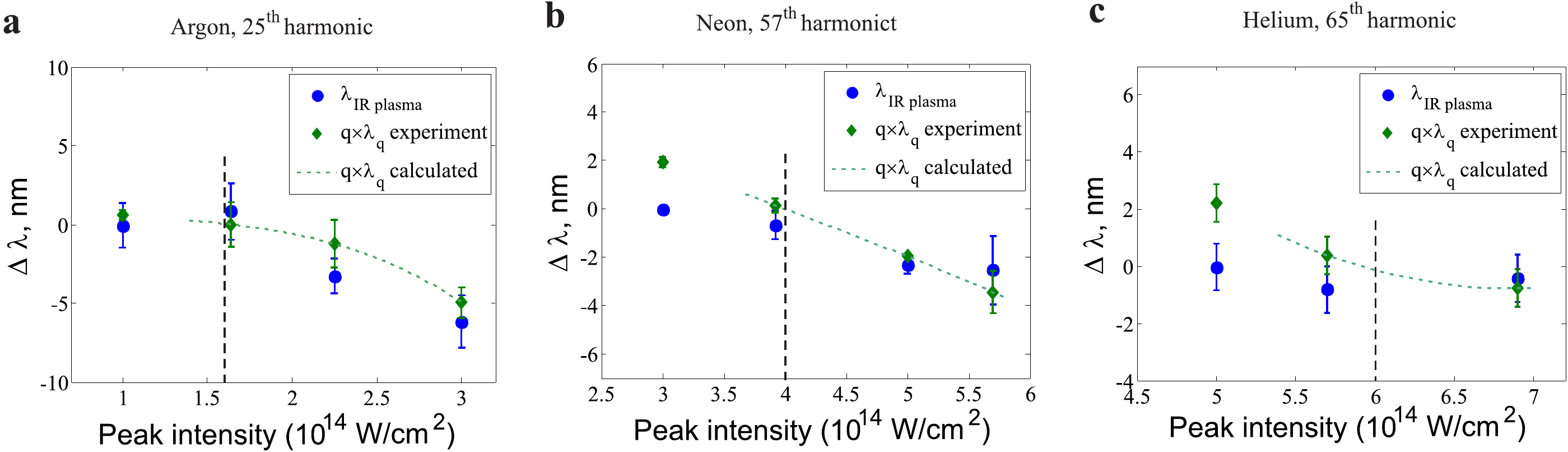}
         \caption{Comparison between the experimentally measured high-order harmonic spectral shifts (green diamonds), driving lasershifts  (blue circles) and calculated values (light green dashed curve) of expected high-order harmonic shifts in accordance with Eq.~(\ref{lambda_q}) for harmonic orders 25$^{\mathrm{th}}$, 57$^{\mathrm{th}}$, 65$^{\mathrm{th}}$ generated in argon (d), neon (e) and helium (f) gases, respectively. Vertical dashed black lines indicate the intensity level $\hat{I}_{PM}$, at which phase-matching radiation is produced from the laser pulse peak.}
        \label{shifts_experiment}
\end{center}
\end{figure}

For the determination of the IR spectral shift, $\Delta\lambda_{\mathrm{IR}}$, we use the recorded laser pulse spectra. This allows us to separate out the effect of the IR laser wavelength shift due to plasma in the resulting high-order harmonic wavelength shift. The driving pulse spectral shifts after the laser-target interaction during HHG in argon, neon and helium are shown in Figs.~\ref{shifts_experiment}(a)-(c), respectively. Besides, the high-order harmonic spectral shifts for the 25$^{\mathrm{th}}$ harmonic in argon, the 57$^{\mathrm{th}}$ in neon and the 65$^{\mathrm{th}}$ in helium are also demonstrated. The spectral shift of the IR is defined as the difference between the central wavelength, obtained as the median value of the spectral intensity distribution of the incoming driving IR laser field, and the one after the interaction with the gas target. A similar procedure is employed for the calculation of the high-order harmonics wavelength shift. 

In order to figure out, which electron trajectory dominates generation process in our experimental conditions, we calculate the expected high-order harmonic spectral shift taking into account the atomic dipole phase according to Eq.~(\ref{total_phase}) for each kind of electron trajectory. In this way, we are able to identify the contributions coming from the long and short trajectories based on the obtained experimental data.
We thus calculate the resulting high-order harmonic spectral shift in wavelength as:
\begin{equation}
\Delta\lambda_q=a \Delta\lambda_{\mathrm{dip},t_1}+(1-a) \Delta\lambda_{\mathrm{dip},t_2}+k\;\frac{\Delta\lambda_{\mathrm{IR}}}{q}.
\label{lambda_q}
\end{equation}
The $a$ factor accounts for the long electron trajectory contribution fraction. In order to find it, we consider values $\Delta\lambda_q$ (measured high harmonic spectral shift), $\Delta\lambda_{\mathrm{IR}}$ (measured driving laser spectral shift), $k$ (correction factor according to Eq.~(\ref{k_factor}) and $q$ (harmonic order) to solve Eq.~(\ref{lambda_q}) for each peak intensity point.
Since the term $-\frac{\partial I}{\partial t}$ in Eq.~(\ref{del_omega_dip}) leads to a zero spectral shift, when the phase-matched harmonic generation is produced at the peak of pulse, it is therefore impossible to determine $a$ for these intensity values. 

Taking into account our XUV spectrometer is not calibrated in absolute terms, we set the zero shift at the peak intensities $\hat{I}_{PM}$ for all the cases (indicated as vertical black dashed lines in the figures. 
\par At low peak intensities, the laser intensity is not enough to reach an ionization level suitable for phase-matching $\eta_{PM}$. Thus, we can not apply Eq.~(\ref{lambda_q}) in order to determine $a$ from experimentally measured data. The harmonics of interest are produced near cutoff at low peak intensities. The $\alpha$ values are known for these intensity values, as can be seen at Fig.~\ref{alpha_plot}. Therefore, it becomes possible to use the measured high harmonic spectral shift in order to determine, at which intensity (and corresponding time) the dominating harmonic generation takes place.
\par We study the correlation of the driving IR laser wavelength shift and that of the 25$^{\mathrm{th}}$ harmonic order in argon as a function of driving laser intensity, as illustrated in Fig.~\ref{shifts_experiment}(a). As can be seen, the obtained high-order harmonic spectral shift $q \Delta\lambda_{q}$ is predominantly given by the fundamental laser wavelength shift $\Delta\lambda_{\mathrm{IR}}$. Even though the uncertainty is high, the $a$ factor is obtained $a \approx0.1$ implying that short electron trajectory dominated the process. 
Thus, the spectral shift induced by the atomic dipole phase is minor for this case and high harmonic spectral shift follows that of driving laser. 
Such a behavior is consistent to the one reported in the literature~\cite{Wahlstrom, kan1995}. 

Next, we consider the spectral shift of the driving laser wavelength and that of the 57$^{\mathrm{th}}$ harmonic order in neon, shown in Fig.~\ref{shifts_experiment}(b). 
Based on the obtained data, by separating the effect of the IR spectral shift, we find that the contribution coming from the atomic dipole phase introduces a redshift at laser peak intensities less than $\hat{I}_{PM}$. At such low laser intensities, plasma does not significantly influence the driving laser pulse. Thus its contribution to the wavelength shift is negligible. As a consequence, we conclude that the spectral shift induced by the atomic dipole phase is responsible for the observed high harmonic spectral shift. The amount of long electron trajectory was found $a\approx0.3$ for HHG in neon. The significant contribution of long electron path can be caused by location of the gas target before laser beam focus, since this position assists phase-matching of the HHG with long electron trajectory \cite{Balcou1997, Chipperfield, Merdji}.

Further, we investigate the correlation between the IR laser wavelength shift and that of the 65$^{\mathrm{th}}$ harmonic order in helium, as demonstrated in Fig.~5(c). A large high-order harmonic redshift $q\Delta\lambda_{q}$ is observed while $\Delta\lambda_{\mathrm{IR}}$ is negligible, before the gas medium becomes significantly ionized. Consequently, for this case we conclude that the contribution from the atomic dipole phase is dominant due to same reason as for case of HHG in neon. 
The amount of long electron trajectory was found $a\approx0.25$ for HHG in helium.
\par During HHG in all cases the origin of harmonic spectral shift for lowest peak intensities was studied. It was found out that cutoff harmonics are mainly generated in the trailing part of the laser pulse peak close to its apex.

\subsection{High-order harmonics spectral broadening}
We studied high-order harmonic spectral broadening as a function of laser intensity. The harmonic spectral width $\Delta_q$ is inversely proportional to high harmonic pulse duration $\Delta \tau_q$ (the FWHM of the $q^\mathrm{th}$-order harmonic intensity), $\Delta_q\propto \frac{1}{\Delta \tau_q}$. 

For our analysis we assume that the harmonic pulse duration is equal to the time interval, where the phase-matched generation occurs, as it is described below. First, we estimate, from the optimal generation conditions, time intervals with $\eta$ suitable for phase-matching. In order to do this, we first obtain the $\eta$-dependence of the length-density product, defined as $L_{\mathrm{coh}} N_a$. 
Combining the calculated $L_{\mathrm{coh}} N_a$ and the experimentally estimated $L N_a$ products, we find phase-matching ionization degree ranges $[\eta_{\mathrm{min}},\eta_{\mathrm{max}}]$ under our optimal experimental conditions. We thus obtain a range between 2.8$\%$ to 4.2$\%$ for $25^{\mathrm{th}}$ harmonic generated in argon, from 0.89 $\%$ to 1.1 $\%$ for $q=57^{\mathrm{th}}$ harmonic generated in neon and from 0.63$\%$ to 0.7$\%$ for $q=65^{\mathrm{th}}$ harmonic order in the case of helium gas. Defining respective phase-matching times $\tau_{PM,\, 1}$ and  $\tau_{PM,\, 2}$ for the lower and upper limits of phase-matching ionization ranges, the corresponding phase-matching time interval $T_{PM}\equiv[\tau_{PM,\, 1}, \, \tau_{PM, \,2}]$ and varies with the laser peak intensity.

We introduce an additional assumption, which takes into account the limitation arising from the single-atom response of the HHG process. This implies that the HHG occurs within a laser intensity region, where the intensity is large enough to produce harmonics above cutoff, i.e.~$I>I_{\mathrm{cutoff}}$. The determination of the laser intensity $I_{\mathrm{cutoff}}$, needed for the generation of a high-order harmonic of energy $\hbar\omega_{\mathrm{cutoff}}$, is done by invoking the semiclassical formula Eq.~(\ref{cutoff})~\cite{maciej1}.

We can then define a time interval, which corresponds to an intensity range fulfilling the condition $I>I_{\mathrm{cutoff}}$. This condition defines the times $\tau_{\mathrm{cutoff,1}}$ and $\tau_{\mathrm{cutoff,2}}$ and consequently we find $T_{\mathrm{cutoff}}\equiv\left[\tau_{\mathrm{cutoff,1}},\,\tau_{\mathrm{cutoff,2}}\right]$.

We now apply both conditions for each particular laser peak intensity and search for the  intersection between the time intervals $T_{PM}$ and $T_{\mathrm{cutoff}}$, so that the final time interval is then $T_q\equiv T_{PM} \cap T_{\mathrm{cutoff}}$. 
At low laser peak intensities, $T_{PM}$ is very large. This happens because the upper boundary of the phase-matching ionization level is not reached. The total high-order harmonic pulse duration $T_q$ is therefore limited by $\tau_{PM,1}$ and $\tau_{\mathrm{cutoff,2}}$. The longest $T_q$ is achieved when $\tau_{PM,2}=\tau_{\mathrm{cutoff,2}}$. As the laser peak intensity increases, the upper-level of the phase matching ionization degree is reached and the duration $T_q$ is fully given by $T_{PM}$, since  $T_{PM}\subseteq T_{\mathrm{cutoff}}$. Thus, the time interval $T_q$ shortens as the laser peak intensity rises. This reduction is caused by the sudden growth of the ionization probability. Following these precepts, we can define the minimal high-order harmonic spectral width, which corresponds to a maximum $T_q$ as $\Delta_{q,\, \mathrm{min}}$. The laser peak intensity, at which this value is reached, is then defined as $I_{\Delta,\, \mathrm{min}}$. 
\begin{figure}[ht]
\begin{center}
\includegraphics[width=1\textwidth]{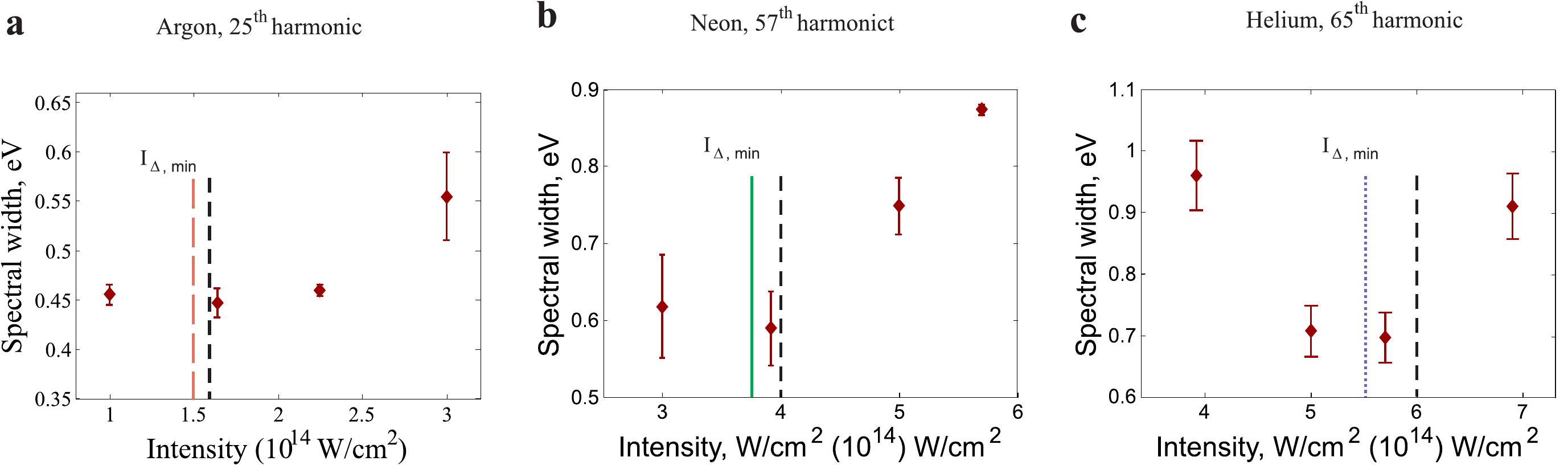}
         \caption{Comparison between the experimentally measured high-order harmonic spectral broadening as a function of the driving laser intensity (red solid diamonds). The presented data are for the 25$^{\mathrm{th}}$, 57$^{\mathrm{th}}$ and 65$^{\mathrm{th}}$ harmonic orders generated in (a) argon, (b) neon and (c) helium, respectively. The vertical long dashed (a), solid (b) and dotted (c) lines correspond to the $I_{\Delta,\, \mathrm{min}}$ values (see the text for details).  The dashed black lines indicate the intensity levels $\hat{I}_{PM}$.}
        \label{figure5}
\end{center}
\end{figure}

We demonstrate the experimentally measured dependence of the high-order harmonic spectral width on the increase of the driving laser intensity in Fig.~5 for the 25$^{\mathrm{th}}$ harmonic order in argon [Fig.~5(a)], the 57$^{\mathrm{th}}$ in neon [Fig.~5(b)] and the 65$^{\mathrm{th}}$ in helium [Fig.~5(c)]. The values $I_{\Delta,\, \mathrm{min}}$ were found to be $1.5\times 10^{14}$ W/cm$^2$, $3.74\times 10^{14}$ W/cm$^2$ and $5.5\times 10^{14}$ W/cm$^2$ for the particular harmonics generated in argon, neon and helium, respectively. As can be seen, the experimentally measured high-order harmonic spectral width shows a minimum around the calculated $I_{\Delta,\, \mathrm{min}}$ values. We thus attribute this spectral width minimum to the maximum high-order harmonic pulse duration $T_q$, obtained considering that the laser intensity is limited, on the one side, by the HHG cutoff, and by the phase-matching conditions, on the other side.
\par We have also studied the influence of the chirp, due to the time-dependent high-order harmonic phase variation, on the high-order harmonic spectral broadening in accordance with assumptions from \cite{chang_phase,Murakami2005}. 
For its calculation we take into account the intensity values in those time intervals, when phase-matched HHG occurs. 
Taking into account both effects, i.e.~the variation of $T_q$ as well as the high-order harmonic chirp due to the atomic dipole phase, we find that the actual effect of the latter is much weaker than the variation of the high-order harmonic pulse duration. Including the chirp contribution into our analysis, its effect is observed only close to regions with a smallest spectral width, i.e. close to $\hat{I}_{PM}$, since second derivative of intensity in time is largest here. In this way, the obtained spectral broadening increases by about 4$\%$ of the maximal spectral width. Thus, it could be assumed negligible under our conditions.

\section{Conclusions}
 
In conclusion, we performed a comprehensive study of the HHG spectral features in noble gases with different ionization potentials. The ability to measure the driving laser's spectral shift makes it possible to disentangle the different contributions to the resulting high-order harmonic wavelength shift. In this way, we can distinguish the wavelength shift coming directly from the driving IR laser pulse from the one produced by the atomic dipole phase. 
We have found the contribution to the resultant high harmonic spectral shift due to atomic dipole phase in case of HHG in argon is negligible. On the other hand, this contribution is non negligible in our experiment in the cases of neon and helium. The relative contributions for our experimental conditions coming from long and short electron trajectories were determined using measured driving laser and high harmonic spectral shifts as well as a computed intensity-dependent $\alpha$ coefficient. Our main observations lead to the conclusion that spectral shift of high-order harmonics corresponds to a shift in the driving laser spectra only when the contributions of the long and short trajectory compensate each other, providing negligible spectral shift due to atomic dipole phase. This is the case as long as the contribution of long trajectory is quite small, about 10$\%$. Otherwise, the spectral shift due to atomic dipole phase is not negligible and resulting high harmonic shift does not follow that of the driving laser. This conclusion is of high importance for experiments, where exact knowledge of the spectral position of high-order harmonics is of high importance.

Additionally, we studied the spectral broadening of the high-order harmonics generated in different gases. The smallest spectral width was determined at the laser peak intensity that provides the largest time-intersection between the microscopic assumptions, when the laser intensity exceeds the cutoff value, and the macroscopic assumptions, when the phase-matching ionization level is achieved. For all atomic species, the narrowest high-order harmonic width was localized at laser peak intensities less than the one corresponding to phase-matching radiation from peak of the pulse (when the medium is not over-ionized). The influence of the atomic dipole chirp on the high-order harmonic spectral broadening was found to be negligible in our study.

The presented data and its analysis provide a significant extension on the knowledge of the HHG spectral characteristics. 

\acknowledgments{We acknowledge the Ti:Sapphire laser system at PALS facility, where the experimental work was carried out with a strong support of J. H\v reb\' i\v cek, T. Med\v r\'ik, and J. Golasowski. We thank R. Jack for helping us with the copyediting of the manuscript.  The results of the  Project LQ1606 were obtained with the financial support of the Ministry of Education, Youth and Sports as part of targeted support from the National Programme of Sustainability II and supported by the project Advanced research using high intensity laser produced photons and particles (CZ.02.1.01/0.0/0.0/16\_019/0000789) from European Regional Development Fund (ADONIS). We also acknowledge project no. CZ.1.07/2.3.00/20.0279 that was co-financed by the European Social Fund and the state budget of the Czech Republic. Operating costs of the PALS facility were covered by the Czech Ministry of Education (Grant LM2015083). The present project was supported by LASERLAB Europe (grant agreement no. 654148, European Union’s Horizon 2020 research and innovation programme); Ministerio de Econom\'{\i}a y Competitividad through Plan Nacional (Grant No. FIS2016-79508-P, FISICATEAMO and Severo Ochoa Excellence Grant No. SEV-2015-0522); funding from the European Unions Horizon 2020 research and Innovation Programme under the Marie Sklodowska-Curie Grant Agreement No. 641272 and Laserlab-Europe (Grant No. EU-H2020 654148), Fundaci\'o Privada Cellex and Generalitat de Catalunya (Grant No.~SGR 874 and CERCA Programme). J. V. is supported by the Grant Agency of the Czech Technical University in Prague, grant No. SGS16/248/OHK4/3T/14. N.S., E.P. and M.L. acknowledge ERC AdG OSYRIS, EU FETPRO QUIC and National Science Centre Poland-Symfonia Grant No. 2016/20/W/ST4/00314.}


\begin{thebibliography}{36}%
\makeatletter
\providecommand \@ifxundefined [1]{%
 \@ifx{#1\undefined}
}%
\providecommand \@ifnum [1]{%
 \ifnum #1\expandafter \@firstoftwo
 \else \expandafter \@secondoftwo
 \fi
}%
\providecommand \@ifx [1]{%
 \ifx #1\expandafter \@firstoftwo
 \else \expandafter \@secondoftwo
 \fi
}%
\providecommand \natexlab [1]{#1}%
\providecommand \enquote  [1]{``#1''}%
\providecommand \bibnamefont  [1]{#1}%
\providecommand \bibfnamefont [1]{#1}%
\providecommand \citenamefont [1]{#1}%
\providecommand \href@noop [0]{\@secondoftwo}%
\providecommand \href [0]{\begingroup \@sanitize@url \@href}%
\providecommand \@href[1]{\@@startlink{#1}\@@href}%
\providecommand \@@href[1]{\endgroup#1\@@endlink}%
\providecommand \@sanitize@url [0]{\catcode `\\12\catcode `\$12\catcode
  `\&12\catcode `\#12\catcode `\^12\catcode `\_12\catcode `\%12\relax}%
\providecommand \@@startlink[1]{}%
\providecommand \@@endlink[0]{}%
\providecommand \url  [0]{\begingroup\@sanitize@url \@url }%
\providecommand \@url [1]{\endgroup\@href {#1}{\urlprefix }}%
\providecommand \urlprefix  [0]{URL }%
\providecommand \Eprint [0]{\href }%
\providecommand \doibase [0]{http://dx.doi.org/}%
\providecommand \selectlanguage [0]{\@gobble}%
\providecommand \bibinfo  [0]{\@secondoftwo}%
\providecommand \bibfield  [0]{\@secondoftwo}%
\providecommand \translation [1]{[#1]}%
\providecommand \BibitemOpen [0]{}%
\providecommand \bibitemStop [0]{}%
\providecommand \bibitemNoStop [0]{.\EOS\space}%
\providecommand \EOS [0]{\spacefactor3000\relax}%
\providecommand \BibitemShut  [1]{\csname bibitem#1\endcsname}%
\let\auto@bib@innerbib\@empty
\bibitem [{\citenamefont {Krausz}\ and\ \citenamefont
  {Ivanov}(2009)}]{ferencreview}%
  \BibitemOpen
  \bibfield  {author} {\bibinfo {author} {\bibfnamefont {F.}~\bibnamefont
  {Krausz}}\ and\ \bibinfo {author} {\bibfnamefont {M.~Y.}\ \bibnamefont
  {Ivanov}},\ }\bibfield  {title} {\enquote {\bibinfo {title} {Attosecond
  physics},}\ }\href@noop {} {\bibfield  {journal} {\bibinfo  {journal} {Rev.
  Mod. Phys.}\ }\textbf {\bibinfo {volume} {81}},\ \bibinfo {pages} {163--234}
  (\bibinfo {year} {2009})}\BibitemShut {NoStop}%
\bibitem [{\citenamefont {Ditmire}\ \emph {et~al.}(1996)\citenamefont
  {Ditmire}, \citenamefont {Gumbrell}, \citenamefont {Smith}, \citenamefont
  {Tisch}, \citenamefont {Meyerhofer},\ and\ \citenamefont
  {Hutchinson}}]{ditmire}%
  \BibitemOpen
  \bibfield  {author} {\bibinfo {author} {\bibfnamefont {T.}~\bibnamefont
  {Ditmire}}, \bibinfo {author} {\bibfnamefont {E.~T.}\ \bibnamefont
  {Gumbrell}}, \bibinfo {author} {\bibfnamefont {R.~A.}\ \bibnamefont {Smith}},
  \bibinfo {author} {\bibfnamefont {J.~W.~G.}\ \bibnamefont {Tisch}}, \bibinfo
  {author} {\bibfnamefont {D.~D.}\ \bibnamefont {Meyerhofer}}, \ and\ \bibinfo
  {author} {\bibfnamefont {M.~H.~R.}\ \bibnamefont {Hutchinson}},\ }\bibfield
  {title} {\enquote {\bibinfo {title} {Spatial coherence measurement of soft
  x-ray radiation produced by high order harmonic generation},}\ }\href@noop {}
  {\bibfield  {journal} {\bibinfo  {journal} {Phys. Rev. Lett.}\ }\textbf
  {\bibinfo {volume} {77}},\ \bibinfo {pages} {4756} (\bibinfo {year}
  {1996})}\BibitemShut {NoStop}%
\bibitem [{\citenamefont {Gautier}\ \emph {et~al.}(2008)\citenamefont
  {Gautier}, \citenamefont {Zeitoun}, \citenamefont {Hauri}, \citenamefont
  {Morlens}, \citenamefont {Rey}, \citenamefont {Valentin}, \citenamefont
  {Papalarazou}, \citenamefont {Goddet}, \citenamefont {Sebban}, \citenamefont
  {Burgy}, \citenamefont {Merc\`ere}, \citenamefont {Idir}, \citenamefont
  {Dovillaire}, \citenamefont {Levecq}, \citenamefont {Bucourt}, \citenamefont
  {Fajardo}, \citenamefont {Merdji},\ and\ \citenamefont {Caumes}}]{gautier}%
  \BibitemOpen
  \bibfield  {author} {\bibinfo {author} {\bibfnamefont {J.}~\bibnamefont
  {Gautier}}, \bibinfo {author} {\bibfnamefont {P.}~\bibnamefont {Zeitoun}},
  \bibinfo {author} {\bibfnamefont {C.}~\bibnamefont {Hauri}}, \bibinfo
  {author} {\bibfnamefont {A.-S.}\ \bibnamefont {Morlens}}, \bibinfo {author}
  {\bibfnamefont {G.}~\bibnamefont {Rey}}, \bibinfo {author} {\bibfnamefont
  {C.}~\bibnamefont {Valentin}}, \bibinfo {author} {\bibfnamefont
  {E.}~\bibnamefont {Papalarazou}}, \bibinfo {author} {\bibfnamefont {J.-P.}\
  \bibnamefont {Goddet}}, \bibinfo {author} {\bibfnamefont {S.}~\bibnamefont
  {Sebban}}, \bibinfo {author} {\bibfnamefont {F.}~\bibnamefont {Burgy}},
  \bibinfo {author} {\bibfnamefont {P.}~\bibnamefont {Merc\`ere}}, \bibinfo
  {author} {\bibfnamefont {M.}~\bibnamefont {Idir}}, \bibinfo {author}
  {\bibfnamefont {G.}~\bibnamefont {Dovillaire}}, \bibinfo {author}
  {\bibfnamefont {X.}~\bibnamefont {Levecq}}, \bibinfo {author} {\bibfnamefont
  {S.}~\bibnamefont {Bucourt}}, \bibinfo {author} {\bibfnamefont
  {M.}~\bibnamefont {Fajardo}}, \bibinfo {author} {\bibfnamefont
  {H.}~\bibnamefont {Merdji}}, \ and\ \bibinfo {author} {\bibfnamefont {J.-P.}\
  \bibnamefont {Caumes}},\ }\bibfield  {title} {\enquote {\bibinfo {title}
  {Optimization of the wave front of high-order harmonics},}\ }\href@noop {}
  {\bibfield  {journal} {\bibinfo  {journal} {Eur. Phys. J. D}\ }\textbf
  {\bibinfo {volume} {48}},\ \bibinfo {pages} {459} (\bibinfo {year}
  {2008})}\BibitemShut {NoStop}%
\bibitem [{\citenamefont {Fleischer}\ \emph {et~al.}(2014)\citenamefont
  {Fleischer}, \citenamefont {Kfier}, \citenamefont {Diskin},\ and\
  \citenamefont {Sidorenko}}]{fleischer}%
  \BibitemOpen
  \bibfield  {author} {\bibinfo {author} {\bibfnamefont {A.}~\bibnamefont
  {Fleischer}}, \bibinfo {author} {\bibfnamefont {O.}~\bibnamefont {Kfier}},
  \bibinfo {author} {\bibfnamefont {T.}~\bibnamefont {Diskin}}, \ and\ \bibinfo
  {author} {\bibfnamefont {P.}~\bibnamefont {Sidorenko}},\ }\bibfield  {title}
  {\enquote {\bibinfo {title} {Spin angular momentum and tunable polarization
  in high-harmonic generation},}\ }\href@noop {} {\bibfield  {journal}
  {\bibinfo  {journal} {Nat. Phot.}\ }\textbf {\bibinfo {volume} {8}},\
  \bibinfo {pages} {543} (\bibinfo {year} {2014})}\BibitemShut {NoStop}%
\bibitem [{\citenamefont {Zhavoronkov}\ and\ \citenamefont
  {Ivanov}(2017)}]{mishaOptLet2017}%
  \BibitemOpen
  \bibfield  {author} {\bibinfo {author} {\bibfnamefont {N.}~\bibnamefont
  {Zhavoronkov}}\ and\ \bibinfo {author} {\bibfnamefont {M.}~\bibnamefont
  {Ivanov}},\ }\bibfield  {title} {\enquote {\bibinfo {title} {Extended
  ellipticity control for attosecond pulses by high harmonic generation},}\
  }\href@noop {} {\bibfield  {journal} {\bibinfo  {journal} {Opt. Lett.}\
  }\textbf {\bibinfo {volume} {42}},\ \bibinfo {pages} {4720} (\bibinfo {year}
  {2017})}\BibitemShut {NoStop}%
\bibitem [{\citenamefont {Takahashi}\ \emph {et~al.}(2007)\citenamefont
  {Takahashi}, \citenamefont {Kanai}, \citenamefont {Ishikawa}, \citenamefont
  {Nabekawa},\ and\ \citenamefont {Midorikawa}}]{taka}%
  \BibitemOpen
  \bibfield  {author} {\bibinfo {author} {\bibfnamefont {E.~J.}\ \bibnamefont
  {Takahashi}}, \bibinfo {author} {\bibfnamefont {T.}~\bibnamefont {Kanai}},
  \bibinfo {author} {\bibfnamefont {K.~L.}\ \bibnamefont {Ishikawa}}, \bibinfo
  {author} {\bibfnamefont {Y.}~\bibnamefont {Nabekawa}}, \ and\ \bibinfo
  {author} {\bibfnamefont {K.}~\bibnamefont {Midorikawa}},\ }\bibfield  {title}
  {\enquote {\bibinfo {title} {Dramatic enhancement of high-order harmonic
  generation},}\ }\href@noop {} {\bibfield  {journal} {\bibinfo  {journal}
  {Phys. Rev. Lett.}\ }\textbf {\bibinfo {volume} {99}},\ \bibinfo {pages}
  {053904} (\bibinfo {year} {2007})}\BibitemShut {NoStop}%
\bibitem [{\citenamefont {Paul}\ \emph {et~al.}(2001)\citenamefont {Paul},
  \citenamefont {Toma}, \citenamefont {Breger}, \citenamefont {Mullot},
  \citenamefont {Auge}, \citenamefont {Balcou}, \citenamefont {Muller},\ and\
  \citenamefont {Agostini}}]{paul}%
  \BibitemOpen
  \bibfield  {author} {\bibinfo {author} {\bibfnamefont {P.}~\bibnamefont
  {Paul}}, \bibinfo {author} {\bibfnamefont {E.}~\bibnamefont {Toma}}, \bibinfo
  {author} {\bibfnamefont {P.}~\bibnamefont {Breger}}, \bibinfo {author}
  {\bibfnamefont {G.}~\bibnamefont {Mullot}}, \bibinfo {author} {\bibfnamefont
  {F.}~\bibnamefont {Auge}}, \bibinfo {author} {\bibfnamefont {Ph.}\
  \bibnamefont {Balcou}}, \bibinfo {author} {\bibfnamefont {H.}~\bibnamefont
  {Muller}}, \ and\ \bibinfo {author} {\bibfnamefont {P.}~\bibnamefont
  {Agostini}},\ }\bibfield  {title} {\enquote {\bibinfo {title} {Observation of
  a train of attosecond pulses from high harmonic generation},}\ }\href@noop {}
  {\bibfield  {journal} {\bibinfo  {journal} {Science}\ }\textbf {\bibinfo
  {volume} {292}},\ \bibinfo {pages} {1689--1692} (\bibinfo {year}
  {2001})}\BibitemShut {NoStop}%
\bibitem [{\citenamefont {Li}\ \emph {et~al.}(2017)\citenamefont {Li},
  \citenamefont {Ren}, \citenamefont {Yin}, \citenamefont {Zhao}, \citenamefont
  {Chew}, \citenamefont {Cheng}, \citenamefont {Cunningham}, \citenamefont
  {Wang}, \citenamefont {Hu}, \citenamefont {Wu}, \citenamefont {Chini},\ and\
  \citenamefont {Chang}}]{chang}%
  \BibitemOpen
  \bibfield  {author} {\bibinfo {author} {\bibfnamefont {J.}~\bibnamefont
  {Li}}, \bibinfo {author} {\bibfnamefont {X.}~\bibnamefont {Ren}}, \bibinfo
  {author} {\bibfnamefont {Y.}~\bibnamefont {Yin}}, \bibinfo {author}
  {\bibfnamefont {K.}~\bibnamefont {Zhao}}, \bibinfo {author} {\bibfnamefont
  {A.}~\bibnamefont {Chew}}, \bibinfo {author} {\bibfnamefont {Y.}~\bibnamefont
  {Cheng}}, \bibinfo {author} {\bibfnamefont {E.}~\bibnamefont {Cunningham}},
  \bibinfo {author} {\bibfnamefont {Y.}~\bibnamefont {Wang}}, \bibinfo {author}
  {\bibfnamefont {S.}~\bibnamefont {Hu}}, \bibinfo {author} {\bibfnamefont
  {Y.}~\bibnamefont {Wu}}, \bibinfo {author} {\bibfnamefont {M.}~\bibnamefont
  {Chini}}, \ and\ \bibinfo {author} {\bibfnamefont {Z.}~\bibnamefont
  {Chang}},\ }\bibfield  {title} {\enquote {\bibinfo {title} {53-attosecond
  x-ray pulses reach the carbon k-edge},}\ }\href@noop {} {\bibfield  {journal}
  {\bibinfo  {journal} {Nat. Commun.}\ }\textbf {\bibinfo {volume} {8}},\
  \bibinfo {pages} {186} (\bibinfo {year} {2017})}\BibitemShut {NoStop}%
\bibitem [{\citenamefont {Corkum}\ and\ \citenamefont
  {Krausz}(2007)}]{Corkum_Nature}%
  \BibitemOpen
  \bibfield  {author} {\bibinfo {author} {\bibfnamefont {P.~B.}\ \bibnamefont
  {Corkum}}\ and\ \bibinfo {author} {\bibfnamefont {F.}~\bibnamefont
  {Krausz}},\ }\bibfield  {title} {\enquote {\bibinfo {title} {Attosecond
  science},}\ }\href@noop {} {\bibfield  {journal} {\bibinfo  {journal} {Nature
  Physics}\ }\textbf {\bibinfo {volume} {3}},\ \bibinfo {pages} {381} (\bibinfo
  {year} {2007})}\BibitemShut {NoStop}%
\bibitem [{\citenamefont {Uiberacker}\ \emph {et~al.}(2007)\citenamefont
  {Uiberacker}, \citenamefont {Uphues}, \citenamefont {Schultze}, \citenamefont
  {Verhoef}, \citenamefont {Yakovlev}, \citenamefont {Kling}, \citenamefont
  {Rauschenberger}, \citenamefont {Kabachnik}, \citenamefont {Schr\"oder},
  \citenamefont {Lezius}, \citenamefont {Kompa}, \citenamefont {Muller},
  \citenamefont {Vrakking}, \citenamefont {Hendel}, \citenamefont {Kleineberg},
  \citenamefont {Heinzmann}, \citenamefont {Drescher},\ and\ \citenamefont
  {Krausz}}]{tunnel}%
  \BibitemOpen
  \bibfield  {author} {\bibinfo {author} {\bibfnamefont {M.}~\bibnamefont
  {Uiberacker}}, \bibinfo {author} {\bibfnamefont {Th.}\ \bibnamefont
  {Uphues}}, \bibinfo {author} {\bibfnamefont {M.}~\bibnamefont {Schultze}},
  \bibinfo {author} {\bibfnamefont {A.~J.}\ \bibnamefont {Verhoef}}, \bibinfo
  {author} {\bibfnamefont {V.}~\bibnamefont {Yakovlev}}, \bibinfo {author}
  {\bibfnamefont {M.~F.}\ \bibnamefont {Kling}}, \bibinfo {author}
  {\bibfnamefont {J.}~\bibnamefont {Rauschenberger}}, \bibinfo {author}
  {\bibfnamefont {N.~M.}\ \bibnamefont {Kabachnik}}, \bibinfo {author}
  {\bibfnamefont {H.}~\bibnamefont {Schr\"oder}}, \bibinfo {author}
  {\bibfnamefont {M.}~\bibnamefont {Lezius}}, \bibinfo {author} {\bibfnamefont
  {K.~L.}\ \bibnamefont {Kompa}}, \bibinfo {author} {\bibfnamefont {H.-G.}\
  \bibnamefont {Muller}}, \bibinfo {author} {\bibfnamefont {M.~J.~J.}\
  \bibnamefont {Vrakking}}, \bibinfo {author} {\bibfnamefont {S.}~\bibnamefont
  {Hendel}}, \bibinfo {author} {\bibfnamefont {U.}~\bibnamefont {Kleineberg}},
  \bibinfo {author} {\bibfnamefont {U.}~\bibnamefont {Heinzmann}}, \bibinfo
  {author} {\bibfnamefont {M.}~\bibnamefont {Drescher}}, \ and\ \bibinfo
  {author} {\bibfnamefont {F.}~\bibnamefont {Krausz}},\ }\bibfield  {title}
  {\enquote {\bibinfo {title} {Attosecond real-time observation of electron
  tunnelling in atoms},}\ }\href@noop {} {\bibfield  {journal} {\bibinfo
  {journal} {Nature}\ }\textbf {\bibinfo {volume} {446}},\ \bibinfo {pages}
  {627} (\bibinfo {year} {2007})}\BibitemShut {NoStop}%
\bibitem [{\citenamefont {Vodungbo}\ \emph {et~al.}(2012)\citenamefont
  {Vodungbo}, \citenamefont {Gautier}, \citenamefont {Lambert}, \citenamefont
  {Sardinha}, \citenamefont {Lozano}, \citenamefont {Sebban}, \citenamefont
  {Ducousso}, \citenamefont {Boutu}, \citenamefont {Li}, \citenamefont {Tudu},
  \citenamefont {Tortarolo}, \citenamefont {Hawaldar}, \citenamefont
  {Delaunay}, \citenamefont {L\'opez-Flores}, \citenamefont {Arabski},
  \citenamefont {Boeglin}, \citenamefont {Merdji}, \citenamefont {Zeitoun},\
  and\ \citenamefont {L\"uning}}]{vodun}%
  \BibitemOpen
  \bibfield  {author} {\bibinfo {author} {\bibfnamefont {B.}~\bibnamefont
  {Vodungbo}}, \bibinfo {author} {\bibfnamefont {J.}~\bibnamefont {Gautier}},
  \bibinfo {author} {\bibfnamefont {G.}~\bibnamefont {Lambert}}, \bibinfo
  {author} {\bibfnamefont {A.~Barszczak}\ \bibnamefont {Sardinha}}, \bibinfo
  {author} {\bibfnamefont {M.}~\bibnamefont {Lozano}}, \bibinfo {author}
  {\bibfnamefont {S.}~\bibnamefont {Sebban}}, \bibinfo {author} {\bibfnamefont
  {M.}~\bibnamefont {Ducousso}}, \bibinfo {author} {\bibfnamefont
  {W.}~\bibnamefont {Boutu}}, \bibinfo {author} {\bibfnamefont
  {K.}~\bibnamefont {Li}}, \bibinfo {author} {\bibfnamefont {B.}~\bibnamefont
  {Tudu}}, \bibinfo {author} {\bibfnamefont {M.}~\bibnamefont {Tortarolo}},
  \bibinfo {author} {\bibfnamefont {R.}~\bibnamefont {Hawaldar}}, \bibinfo
  {author} {\bibfnamefont {R.}~\bibnamefont {Delaunay}}, \bibinfo {author}
  {\bibfnamefont {V.}~\bibnamefont {L\'opez-Flores}}, \bibinfo {author}
  {\bibfnamefont {J.}~\bibnamefont {Arabski}}, \bibinfo {author} {\bibfnamefont
  {C.}~\bibnamefont {Boeglin}}, \bibinfo {author} {\bibfnamefont
  {H.}~\bibnamefont {Merdji}}, \bibinfo {author} {\bibfnamefont {Ph.}\
  \bibnamefont {Zeitoun}}, \ and\ \bibinfo {author} {\bibfnamefont
  {J.}~\bibnamefont {L\"uning}},\ }\bibfield  {title} {\enquote {\bibinfo
  {title} {Laser-induced ultrafast demagnetization in the presence of a
  nanoscale magnetic domain network},}\ }\href@noop {} {\bibfield  {journal}
  {\bibinfo  {journal} {Nat. Commun.}\ }\textbf {\bibinfo {volume} {3}},\
  \bibinfo {pages} {999} (\bibinfo {year} {2012})}\BibitemShut {NoStop}%
\bibitem [{\citenamefont {Zhou}\ \emph {et~al.}(2008)\citenamefont {Zhou},
  \citenamefont {Lock}, \citenamefont {Li}, \citenamefont {Wagner},
  \citenamefont {Murnane},\ and\ \citenamefont {Kapteyn}}]{zhou}%
  \BibitemOpen
  \bibfield  {author} {\bibinfo {author} {\bibfnamefont {X.}~\bibnamefont
  {Zhou}}, \bibinfo {author} {\bibfnamefont {R.}~\bibnamefont {Lock}}, \bibinfo
  {author} {\bibfnamefont {W.}~\bibnamefont {Li}}, \bibinfo {author}
  {\bibfnamefont {N.}~\bibnamefont {Wagner}}, \bibinfo {author} {\bibfnamefont
  {M.}~\bibnamefont {Murnane}}, \ and\ \bibinfo {author} {\bibfnamefont
  {H.}~\bibnamefont {Kapteyn}},\ }\bibfield  {title} {\enquote {\bibinfo
  {title} {Molecular recollision interferometry in high harmonic generation},}\
  }\href@noop {} {\bibfield  {journal} {\bibinfo  {journal} {Phys. Rev. Lett.}\
  }\textbf {\bibinfo {volume} {100}},\ \bibinfo {pages} {073902} (\bibinfo
  {year} {2008})}\BibitemShut {NoStop}%
\bibitem [{\citenamefont {Tzallas}\ \emph {et~al.}(2011)\citenamefont
  {Tzallas}, \citenamefont {Skantzakis}, \citenamefont {Nikolopoulos},
  \citenamefont {Tsakiris},\ and\ \citenamefont {Charalambidis}}]{tzallas}%
  \BibitemOpen
  \bibfield  {author} {\bibinfo {author} {\bibfnamefont {P.}~\bibnamefont
  {Tzallas}}, \bibinfo {author} {\bibfnamefont {E.}~\bibnamefont {Skantzakis}},
  \bibinfo {author} {\bibfnamefont {L.~A.~A.}\ \bibnamefont {Nikolopoulos}},
  \bibinfo {author} {\bibfnamefont {G.~D.}\ \bibnamefont {Tsakiris}}, \ and\
  \bibinfo {author} {\bibfnamefont {D.}~\bibnamefont {Charalambidis}},\
  }\bibfield  {title} {\enquote {\bibinfo {title} {Extreme-ultraviolet
  pump-��probe studies of one-femtosecond-scale electron dynamics},}\
  }\href@noop {} {\bibfield  {journal} {\bibinfo  {journal} {Nat. Phys.}\
  }\textbf {\bibinfo {volume} {7}},\ \bibinfo {pages} {781--784} (\bibinfo
  {year} {2011})}\BibitemShut {NoStop}%
\bibitem [{\citenamefont {Manschwetus}\ \emph {et~al.}(2016)\citenamefont
  {Manschwetus}, \citenamefont {Rading}, \citenamefont {Campi}, \citenamefont
  {Maclot}, \citenamefont {Coudert-Alteirac}, \citenamefont {Lahl},
  \citenamefont {Wikmark}, \citenamefont {Rudawski}, \citenamefont {Heyl},
  \citenamefont {Farkas}, \citenamefont {Mohamed}, \citenamefont {L'Huillier},\
  and\ \citenamefont {Johnsson}}]{twophoton}%
  \BibitemOpen
  \bibfield  {author} {\bibinfo {author} {\bibfnamefont {B.}~\bibnamefont
  {Manschwetus}}, \bibinfo {author} {\bibfnamefont {L.}~\bibnamefont {Rading}},
  \bibinfo {author} {\bibfnamefont {F.}~\bibnamefont {Campi}}, \bibinfo
  {author} {\bibfnamefont {S.}~\bibnamefont {Maclot}}, \bibinfo {author}
  {\bibfnamefont {H.}~\bibnamefont {Coudert-Alteirac}}, \bibinfo {author}
  {\bibfnamefont {J.}~\bibnamefont {Lahl}}, \bibinfo {author} {\bibfnamefont
  {H.}~\bibnamefont {Wikmark}}, \bibinfo {author} {\bibfnamefont
  {P.}~\bibnamefont {Rudawski}}, \bibinfo {author} {\bibfnamefont {C.~M.}\
  \bibnamefont {Heyl}}, \bibinfo {author} {\bibfnamefont {B.}~\bibnamefont
  {Farkas}}, \bibinfo {author} {\bibfnamefont {T.}~\bibnamefont {Mohamed}},
  \bibinfo {author} {\bibfnamefont {A.}~\bibnamefont {L'Huillier}}, \ and\
  \bibinfo {author} {\bibfnamefont {P.}~\bibnamefont {Johnsson}},\ }\bibfield
  {title} {\enquote {\bibinfo {title} {Two-photon double ionization of neon
  using an intense attosecond pulse train},}\ }\href@noop {} {\bibfield
  {journal} {\bibinfo  {journal} {Phys. Rev. A}\ }\textbf {\bibinfo {volume}
  {93}},\ \bibinfo {pages} {061402(R)} (\bibinfo {year} {2016})}\BibitemShut
  {NoStop}%
\bibitem [{\citenamefont {Corkum}(1993)}]{corkum}%
  \BibitemOpen
  \bibfield  {author} {\bibinfo {author} {\bibfnamefont {P.~B.}\ \bibnamefont
  {Corkum}},\ }\bibfield  {title} {\enquote {\bibinfo {title} {Plasma
  perspective on strong field multiphoton ionization},}\ }\href@noop {}
  {\bibfield  {journal} {\bibinfo  {journal} {Phys. Rev. Lett.}\ }\textbf
  {\bibinfo {volume} {71}},\ \bibinfo {pages} {1994} (\bibinfo {year}
  {1993})}\BibitemShut {NoStop}%
\bibitem [{\citenamefont {Kulander}\ \emph {et~al.}(1993)\citenamefont
  {Kulander}, \citenamefont {Schafer},\ and\ \citenamefont
  {Krause}}]{kulandernato}%
  \BibitemOpen
  \bibfield  {author} {\bibinfo {author} {\bibfnamefont {K.~C.}\ \bibnamefont
  {Kulander}}, \bibinfo {author} {\bibfnamefont {K.~J.}\ \bibnamefont
  {Schafer}}, \ and\ \bibinfo {author} {\bibfnamefont {K.~L.}\ \bibnamefont
  {Krause}},\ }\bibfield  {title} {\enquote {\bibinfo {title} {Super intense
  atom physics},}\ }in\ \href@noop {} {\emph {\bibinfo {booktitle} {Vol. 316 of
  NATO Advanced Study Institute Series B: Physics}}},\ \bibinfo {editor}
  {edited by\ \bibinfo {editor} {\bibfnamefont {B.}~\bibnamefont {Piraux}},
  \bibinfo {editor} {\bibfnamefont {A.}~\bibnamefont {L'Huillier}}, \ and\
  \bibinfo {editor} {\bibfnamefont {K.}~\bibnamefont {Rzazewski}}}\ (\bibinfo
  {publisher} {Plenum},\ \bibinfo {address} {New York},\ \bibinfo {year}
  {1993})\ p.~\bibinfo {pages} {95}\BibitemShut {NoStop}%
\bibitem [{\citenamefont {Krause}\ \emph {et~al.}(1992)\citenamefont {Krause},
  \citenamefont {Schafer},\ and\ \citenamefont {Kulander}}]{kulander1992}%
  \BibitemOpen
  \bibfield  {author} {\bibinfo {author} {\bibfnamefont {J.~L.}\ \bibnamefont
  {Krause}}, \bibinfo {author} {\bibfnamefont {K.~J.}\ \bibnamefont {Schafer}},
  \ and\ \bibinfo {author} {\bibfnamefont {K.~C.}\ \bibnamefont {Kulander}},\
  }\bibfield  {title} {\enquote {\bibinfo {title} {High-order harmonic
  generation from atoms and ions in the high intensity regime},}\ }\href@noop
  {} {\bibfield  {journal} {\bibinfo  {journal} {Phys. Rev. Lett.}\ }\textbf
  {\bibinfo {volume} {68}},\ \bibinfo {pages} {3535} (\bibinfo {year}
  {1992})}\BibitemShut {NoStop}%
\bibitem [{\citenamefont {Ruchon}\ \emph {et~al.}(2008)\citenamefont {Ruchon},
  \citenamefont {Hauri}, \citenamefont {Varj\'u}, \citenamefont {Gustafsson},
  \citenamefont {L\'opez-Martens},\ and\ \citenamefont {L'Huillier}}]{Ruchon}%
  \BibitemOpen
  \bibfield  {author} {\bibinfo {author} {\bibfnamefont {T.}~\bibnamefont
  {Ruchon}}, \bibinfo {author} {\bibfnamefont {C.P.}\ \bibnamefont {Hauri}},
  \bibinfo {author} {\bibfnamefont {K.}~\bibnamefont {Varj\'u}}, \bibinfo
  {author} {\bibfnamefont {E.}~\bibnamefont {Gustafsson}}, \bibinfo {author}
  {\bibfnamefont {R.}~\bibnamefont {L\'opez-Martens}}, \ and\ \bibinfo {author}
  {\bibfnamefont {A.}~\bibnamefont {L'Huillier}},\ }\bibfield  {title}
  {\enquote {\bibinfo {title} {Macroscopic effects in attosecond pulse
  generation},}\ }\href@noop {} {\bibfield  {journal} {\bibinfo  {journal} {New
  J. Phys.}\ }\textbf {\bibinfo {volume} {10}},\ \bibinfo {pages} {025027}
  (\bibinfo {year} {2008})}\BibitemShut {NoStop}%
\bibitem [{\citenamefont {Delfin}\ \emph {et~al.}(1999)\citenamefont {Delfin},
  \citenamefont {Altucci}, \citenamefont {Filippo}, \citenamefont {de~Lisio},
  \citenamefont {Gaarde}, \citenamefont {L'Huillier}, \citenamefont {Roos},\
  and\ \citenamefont {Wahlstr\"om}}]{Delfin}%
  \BibitemOpen
  \bibfield  {author} {\bibinfo {author} {\bibfnamefont {C.}~\bibnamefont
  {Delfin}}, \bibinfo {author} {\bibfnamefont {C.}~\bibnamefont {Altucci}},
  \bibinfo {author} {\bibfnamefont {F.~De}\ \bibnamefont {Filippo}}, \bibinfo
  {author} {\bibfnamefont {C.}~\bibnamefont {de~Lisio}}, \bibinfo {author}
  {\bibfnamefont {M.B.}\ \bibnamefont {Gaarde}}, \bibinfo {author}
  {\bibfnamefont {A.}~\bibnamefont {L'Huillier}}, \bibinfo {author}
  {\bibfnamefont {L.}~\bibnamefont {Roos}}, \ and\ \bibinfo {author}
  {\bibfnamefont {C.-G.}\ \bibnamefont {Wahlstr\"om}},\ }\bibfield  {title}
  {\enquote {\bibinfo {title} {Influence of the medium length on high-order
  harmonic generation},}\ }\href@noop {} {\bibfield  {journal} {\bibinfo
  {journal} {J. Phys. B}\ }\textbf {\bibinfo {volume} {32}},\ \bibinfo {pages}
  {5397} (\bibinfo {year} {1999})}\BibitemShut {NoStop}%
\bibitem [{\citenamefont {Merdji}\ \emph {et~al.}(2006)\citenamefont {Merdji},
  \citenamefont {Kova\u{c}ev}, \citenamefont {Boutu}, \citenamefont
  {Sali\`eres}, \citenamefont {Vernay},\ and\ \citenamefont
  {Carr\'e}}]{Merdji}%
  \BibitemOpen
  \bibfield  {author} {\bibinfo {author} {\bibfnamefont {H.}~\bibnamefont
  {Merdji}}, \bibinfo {author} {\bibfnamefont {M.}~\bibnamefont {Kova\u{c}ev}},
  \bibinfo {author} {\bibfnamefont {W.}~\bibnamefont {Boutu}}, \bibinfo
  {author} {\bibfnamefont {P.}~\bibnamefont {Sali\`eres}}, \bibinfo {author}
  {\bibfnamefont {F.}~\bibnamefont {Vernay}}, \ and\ \bibinfo {author}
  {\bibfnamefont {B.}~\bibnamefont {Carr\'e}},\ }\bibfield  {title} {\enquote
  {\bibinfo {title} {Macroscopic control of high-order harmonics quantum-path
  components for the generation of attosecond pulses},}\ }\href@noop {}
  {\bibfield  {journal} {\bibinfo  {journal} {Phys. Rev. A}\ }\textbf {\bibinfo
  {volume} {74}},\ \bibinfo {pages} {043804} (\bibinfo {year}
  {2006})}\BibitemShut {NoStop}%
\bibitem [{\citenamefont {Lewenstein}\ \emph {et~al.}(1994)\citenamefont
  {Lewenstein}, \citenamefont {Balcou}, \citenamefont {Ivanov}, \citenamefont
  {L'Huillier},\ and\ \citenamefont {Corkum}}]{maciej1}%
  \BibitemOpen
  \bibfield  {author} {\bibinfo {author} {\bibfnamefont {M.}~\bibnamefont
  {Lewenstein}}, \bibinfo {author} {\bibfnamefont {Ph.}\ \bibnamefont
  {Balcou}}, \bibinfo {author} {\bibfnamefont {M.~Y.}\ \bibnamefont {Ivanov}},
  \bibinfo {author} {\bibfnamefont {A.}~\bibnamefont {L'Huillier}}, \ and\
  \bibinfo {author} {\bibfnamefont {P.~B.}\ \bibnamefont {Corkum}},\ }\bibfield
   {title} {\enquote {\bibinfo {title} {Theory of high-harmonic generation by
  low-frequency laser fields},}\ }\href@noop {} {\bibfield  {journal} {\bibinfo
   {journal} {Phys. Rev. A}\ }\textbf {\bibinfo {volume} {49}},\ \bibinfo
  {pages} {2117--2132} (\bibinfo {year} {1994})}\BibitemShut {NoStop}%
\bibitem [{\citenamefont {Wahlstr\"om}\ \emph {et~al.}(1993)\citenamefont
  {Wahlstr\"om}, \citenamefont {Larsson}, \citenamefont {Persson},
  \citenamefont {Starczewski}, \citenamefont {Svanberg}, \citenamefont
  {Sali\`eres}, \citenamefont {Balcou},\ and\ \citenamefont
  {L'Huillier}}]{Wahlstrom}%
  \BibitemOpen
  \bibfield  {author} {\bibinfo {author} {\bibfnamefont {C.-G.}\ \bibnamefont
  {Wahlstr\"om}}, \bibinfo {author} {\bibfnamefont {J.}~\bibnamefont
  {Larsson}}, \bibinfo {author} {\bibfnamefont {A.}~\bibnamefont {Persson}},
  \bibinfo {author} {\bibfnamefont {T.}~\bibnamefont {Starczewski}}, \bibinfo
  {author} {\bibfnamefont {S.}~\bibnamefont {Svanberg}}, \bibinfo {author}
  {\bibfnamefont {P.}~\bibnamefont {Sali\`eres}}, \bibinfo {author}
  {\bibfnamefont {Ph.}\ \bibnamefont {Balcou}}, \ and\ \bibinfo {author}
  {\bibfnamefont {A.}~\bibnamefont {L'Huillier}},\ }\bibfield  {title}
  {\enquote {\bibinfo {title} {High-order harmonic generation in rare gases
  with an intense short-pulse laser},}\ }\href@noop {} {\bibfield  {journal}
  {\bibinfo  {journal} {Phys. Rev. A}\ }\textbf {\bibinfo {volume} {48}},\
  \bibinfo {pages} {4709} (\bibinfo {year} {1993})}\BibitemShut {NoStop}%
\bibitem [{\citenamefont {Rae}\ and\ \citenamefont {Burnett}(1993)}]{Rae93}%
  \BibitemOpen
  \bibfield  {author} {\bibinfo {author} {\bibfnamefont {S.~C.}\ \bibnamefont
  {Rae}}\ and\ \bibinfo {author} {\bibfnamefont {K.}~\bibnamefont {Burnett}},\
  }\bibfield  {title} {\enquote {\bibinfo {title} {Harmonic generation and
  phase matching in the tunnelling limit},}\ }\href@noop {} {\bibfield
  {journal} {\bibinfo  {journal} {J. Phys. B}\ }\textbf {\bibinfo {volume}
  {26}},\ \bibinfo {pages} {1509} (\bibinfo {year} {1993})}\BibitemShut
  {NoStop}%
\bibitem [{\citenamefont {Chang}(2011)}]{chang2011}%
  \BibitemOpen
  \bibfield  {author} {\bibinfo {author} {\bibfnamefont {Z.}~\bibnamefont
  {Chang}},\ }\href@noop {} {\emph {\bibinfo {title} {Fundamentals of
  Attosecond Optics}}}\ (\bibinfo  {publisher} {CRC Press},\ \bibinfo {address}
  {Boca Raton, USA},\ \bibinfo {year} {2011})\BibitemShut {NoStop}%
\bibitem [{\citenamefont {Blanc}\ \emph {et~al.}(1993)\citenamefont {Blanc},
  \citenamefont {Sauerbrey}, \citenamefont {Rae},\ and\ \citenamefont
  {Burnett}}]{leblanc}%
  \BibitemOpen
  \bibfield  {author} {\bibinfo {author} {\bibfnamefont {S.~P.~Le}\
  \bibnamefont {Blanc}}, \bibinfo {author} {\bibfnamefont {R.}~\bibnamefont
  {Sauerbrey}}, \bibinfo {author} {\bibfnamefont {S.~C.}\ \bibnamefont {Rae}},
  \ and\ \bibinfo {author} {\bibfnamefont {K.}~\bibnamefont {Burnett}},\
  }\bibfield  {title} {\enquote {\bibinfo {title} {Spectral blue shifting of a
  femtosecond laser pulse propagating through a high-pressure gas},}\
  }\href@noop {} {\bibfield  {journal} {\bibinfo  {journal} {J. Opt. Soc. Am.
  B}\ }\textbf {\bibinfo {volume} {10}},\ \bibinfo {pages} {1801--1809}
  (\bibinfo {year} {1993})}\BibitemShut {NoStop}%
\bibitem [{\citenamefont {Constant}\ \emph {et~al.}(1999)\citenamefont
  {Constant}, \citenamefont {Garzella}, \citenamefont {Breger}, \citenamefont
  {M\'evel}, \citenamefont {Dorrer}, \citenamefont {Blanc}, \citenamefont
  {Salin},\ and\ \citenamefont {Agostini}}]{Constant}%
  \BibitemOpen
  \bibfield  {author} {\bibinfo {author} {\bibfnamefont {E.}~\bibnamefont
  {Constant}}, \bibinfo {author} {\bibfnamefont {D.}~\bibnamefont {Garzella}},
  \bibinfo {author} {\bibfnamefont {P.}~\bibnamefont {Breger}}, \bibinfo
  {author} {\bibfnamefont {E.}~\bibnamefont {M\'evel}}, \bibinfo {author}
  {\bibfnamefont {C.}~\bibnamefont {Dorrer}}, \bibinfo {author} {\bibfnamefont
  {C.~Le}\ \bibnamefont {Blanc}}, \bibinfo {author} {\bibfnamefont
  {F.}~\bibnamefont {Salin}}, \ and\ \bibinfo {author} {\bibfnamefont
  {P.}~\bibnamefont {Agostini}},\ }\bibfield  {title} {\enquote {\bibinfo
  {title} {Optimizing high harmonic generation in absorbing gases: Model and
  experiment},}\ }\href@noop {} {\bibfield  {journal} {\bibinfo  {journal}
  {Phys. Rev. Lett.}\ }\textbf {\bibinfo {volume} {82}},\ \bibinfo {pages}
  {1668} (\bibinfo {year} {1999})}\BibitemShut {NoStop}%
\bibitem [{\citenamefont {Lewenstein}\ \emph {et~al.}(1995)\citenamefont
  {Lewenstein}, \citenamefont {Sali\`eres},\ and\ \citenamefont
  {L'Huillier}}]{Lewenstein1}%
  \BibitemOpen
  \bibfield  {author} {\bibinfo {author} {\bibfnamefont {M.}~\bibnamefont
  {Lewenstein}}, \bibinfo {author} {\bibfnamefont {P.}~\bibnamefont
  {Sali\`eres}}, \ and\ \bibinfo {author} {\bibfnamefont {A.}~\bibnamefont
  {L'Huillier}},\ }\bibfield  {title} {\enquote {\bibinfo {title} {Phase of the
  atomic polarization in high-order harmonic generation},}\ }\href@noop {}
  {\bibfield  {journal} {\bibinfo  {journal} {Phys. Rev. A}\ }\textbf {\bibinfo
  {volume} {52}},\ \bibinfo {pages} {4747} (\bibinfo {year}
  {1995})}\BibitemShut {NoStop}%
\bibitem [{\citenamefont {Ivanov}\ and\ \citenamefont
  {Smirnova}(2014)}]{HHGTutorial}%
  \BibitemOpen
  \bibfield  {author} {\bibinfo {author} {\bibfnamefont {M.}~\bibnamefont
  {Ivanov}}\ and\ \bibinfo {author} {\bibfnamefont {O.}~\bibnamefont
  {Smirnova}},\ }\bibfield  {title} {\enquote {\bibinfo {title} {Multielectron
  high harmonic generation: simple man on a complex plane},}\ }in\ \href@noop
  {} {\emph {\bibinfo {booktitle} {Attosecond and {XUV} Physics: Ultrafast
  Dynamics and Spectroscopy}}},\ \bibinfo {editor} {edited by\ \bibinfo
  {editor} {\bibfnamefont {T}~\bibnamefont {Schultz}}\ and\ \bibinfo {editor}
  {\bibfnamefont {M}~\bibnamefont {Vrakking}}}\ (\bibinfo  {publisher}
  {Wiley-{VCH}},\ \bibinfo {address} {Weinheim},\ \bibinfo {year} {2014})\ pp.\
  \bibinfo {pages} {201--256},\ \Eprint
  {arXiv:1304.2413/http://arxiv.org/abs/1304.2413}
  {http://arxiv.org/abs/1304.2413} \BibitemShut {NoStop}%
\bibitem [{\citenamefont {Catoire}\ \emph {et~al.}(2016)\citenamefont
  {Catoire}, \citenamefont {Ferr\'e}, \citenamefont {Hort}, \citenamefont
  {Dubrouil}, \citenamefont {Quintard}, \citenamefont {Descamps}, \citenamefont
  {Petit}, \citenamefont {Burgy}, \citenamefont {M\'avel}, \citenamefont
  {Mairesse},\ and\ \citenamefont {Constant}}]{Catoire}%
  \BibitemOpen
  \bibfield  {author} {\bibinfo {author} {\bibfnamefont {F.}~\bibnamefont
  {Catoire}}, \bibinfo {author} {\bibfnamefont {A.}~\bibnamefont {Ferr\'e}},
  \bibinfo {author} {\bibfnamefont {O.}~\bibnamefont {Hort}}, \bibinfo {author}
  {\bibfnamefont {A.}~\bibnamefont {Dubrouil}}, \bibinfo {author}
  {\bibfnamefont {L.}~\bibnamefont {Quintard}}, \bibinfo {author}
  {\bibfnamefont {D.}~\bibnamefont {Descamps}}, \bibinfo {author}
  {\bibfnamefont {S.}~\bibnamefont {Petit}}, \bibinfo {author} {\bibfnamefont
  {F.}~\bibnamefont {Burgy}}, \bibinfo {author} {\bibfnamefont
  {E.}~\bibnamefont {M\'avel}}, \bibinfo {author} {\bibfnamefont
  {Y.}~\bibnamefont {Mairesse}}, \ and\ \bibinfo {author} {\bibfnamefont
  {E.}~\bibnamefont {Constant}},\ }\bibfield  {title} {\enquote {\bibinfo
  {title} {Complex structure of spatially resolved high-order-harmonic
  spectra},}\ }\href@noop {} {\bibfield  {journal} {\bibinfo  {journal} {Phys.
  Rev. A}\ }\textbf {\bibinfo {volume} {94}},\ \bibinfo {pages} {063401}
  (\bibinfo {year} {2016})}\BibitemShut {NoStop}%
\bibitem [{\citenamefont {Tong}\ and\ \citenamefont {Lin}(2005)}]{tonglin}%
  \BibitemOpen
  \bibfield  {author} {\bibinfo {author} {\bibfnamefont {X.~M.}\ \bibnamefont
  {Tong}}\ and\ \bibinfo {author} {\bibfnamefont {C.~D.}\ \bibnamefont {Lin}},\
  }\bibfield  {title} {\enquote {\bibinfo {title} {Empirical formula for static
  field ionization rates of atoms and molecules by lasers in the
  barrier-suppression regime},}\ }\href@noop {} {\bibfield  {journal} {\bibinfo
   {journal} {J. Phys. B}\ }\textbf {\bibinfo {volume} {38}},\ \bibinfo {pages}
  {2593--2600} (\bibinfo {year} {2005})}\BibitemShut {NoStop}%
\bibitem [{\citenamefont {Milo\v{s}evi\'c}\ and\ \citenamefont
  {Becker}(2002)}]{miloandbecker}%
  \BibitemOpen
  \bibfield  {author} {\bibinfo {author} {\bibfnamefont {D.~B.}\ \bibnamefont
  {Milo\v{s}evi\'c}}\ and\ \bibinfo {author} {\bibfnamefont {W.}~\bibnamefont
  {Becker}},\ }\bibfield  {title} {\enquote {\bibinfo {title} {Role of long
  quantum orbits in high-order harmonic generation},}\ }\href@noop {}
  {\bibfield  {journal} {\bibinfo  {journal} {Phys. Rev. A}\ }\textbf {\bibinfo
  {volume} {66}},\ \bibinfo {pages} {063417} (\bibinfo {year}
  {2002})}\BibitemShut {NoStop}%
\bibitem [{\citenamefont {Murakami}\ \emph {et~al.}(2005)\citenamefont
  {Murakami}, \citenamefont {Mauritsson},\ and\ \citenamefont
  {Gaarde}}]{Murakami2005}%
  \BibitemOpen
  \bibfield  {author} {\bibinfo {author} {\bibfnamefont {M.}~\bibnamefont
  {Murakami}}, \bibinfo {author} {\bibfnamefont {J.}~\bibnamefont
  {Mauritsson}}, \ and\ \bibinfo {author} {\bibfnamefont {M.~B.}\ \bibnamefont
  {Gaarde}},\ }\bibfield  {title} {\enquote {\bibinfo {title} {Frequency-chirp
  rates of harmonics driven by a few-cycle pulse},}\ }\href@noop {} {\bibfield
  {journal} {\bibinfo  {journal} {Phys. Rev. A}\ }\textbf {\bibinfo {volume}
  {72}},\ \bibinfo {pages} {023413} (\bibinfo {year} {2005})}\BibitemShut
  {NoStop}%
\bibitem [{\citenamefont {Kan}\ \emph {et~al.}(1995)\citenamefont {Kan},
  \citenamefont {Capjack}, \citenamefont {Rankin},\ and\ \citenamefont
  {Burnett}}]{kan1995}%
  \BibitemOpen
  \bibfield  {author} {\bibinfo {author} {\bibfnamefont {C.}~\bibnamefont
  {Kan}}, \bibinfo {author} {\bibfnamefont {C.~E.}\ \bibnamefont {Capjack}},
  \bibinfo {author} {\bibfnamefont {R.}~\bibnamefont {Rankin}}, \ and\ \bibinfo
  {author} {\bibfnamefont {N.~H}\ \bibnamefont {Burnett}},\ }\bibfield  {title}
  {\enquote {\bibinfo {title} {Spectral and temporal structure in high harmonic
  emission from ionizing atomic gases},}\ }\href@noop {} {\bibfield  {journal}
  {\bibinfo  {journal} {Phys. Rev. A}\ }\textbf {\bibinfo {volume} {52}},\
  \bibinfo {pages} {R4336} (\bibinfo {year} {1995})}\BibitemShut {NoStop}%
\bibitem [{\citenamefont {Balcou}\ \emph {et~al.}(1997)\citenamefont {Balcou},
  \citenamefont {Sali\`eres}, \citenamefont {L'Huillier},\ and\ \citenamefont
  {Lewenstein}}]{Balcou1997}%
  \BibitemOpen
  \bibfield  {author} {\bibinfo {author} {\bibfnamefont {P.}~\bibnamefont
  {Balcou}}, \bibinfo {author} {\bibfnamefont {P.}~\bibnamefont {Sali\`eres}},
  \bibinfo {author} {\bibfnamefont {A.}~\bibnamefont {L'Huillier}}, \ and\
  \bibinfo {author} {\bibfnamefont {M.}~\bibnamefont {Lewenstein}},\ }\bibfield
   {title} {\enquote {\bibinfo {title} {Generalized phase-matching conditions
  for high harmonics: The role of field-gradient forces},}\ }\href@noop {}
  {\bibfield  {journal} {\bibinfo  {journal} {Phys. Rev. A}\ }\textbf {\bibinfo
  {volume} {55}},\ \bibinfo {pages} {3204} (\bibinfo {year}
  {1997})}\BibitemShut {NoStop}%
\bibitem [{\citenamefont {Chipperfield}\ \emph {et~al.}(2006)\citenamefont
  {Chipperfield}, \citenamefont {Knight}, \citenamefont {Tisch},\ and\
  \citenamefont {Marangos}}]{Chipperfield}%
  \BibitemOpen
  \bibfield  {author} {\bibinfo {author} {\bibfnamefont {L.~E.}\ \bibnamefont
  {Chipperfield}}, \bibinfo {author} {\bibfnamefont {P.~L.}\ \bibnamefont
  {Knight}}, \bibinfo {author} {\bibfnamefont {J.~W.~G.}\ \bibnamefont
  {Tisch}}, \ and\ \bibinfo {author} {\bibfnamefont {J.~P.}\ \bibnamefont
  {Marangos}},\ }\bibfield  {title} {\enquote {\bibinfo {title} {Tracking
  individual electron trajectories in a high harmonic spectrum},}\ }\href@noop
  {} {\bibfield  {journal} {\bibinfo  {journal} {Opt. Commun.}\ }\textbf
  {\bibinfo {volume} {264}},\ \bibinfo {pages} {494--501} (\bibinfo {year}
  {2006})}\BibitemShut {NoStop}%
\bibitem [{\citenamefont {Chang}\ \emph {et~al.}(1998)\citenamefont {Chang},
  \citenamefont {Rundquist}, \citenamefont {Wang}, \citenamefont {Christov},
  \citenamefont {Kapteyn},\ and\ \citenamefont {Murnane}}]{chang_phase}%
  \BibitemOpen
  \bibfield  {author} {\bibinfo {author} {\bibfnamefont {Z.}~\bibnamefont
  {Chang}}, \bibinfo {author} {\bibfnamefont {A.}~\bibnamefont {Rundquist}},
  \bibinfo {author} {\bibfnamefont {H.}~\bibnamefont {Wang}}, \bibinfo {author}
  {\bibfnamefont {I.}~\bibnamefont {Christov}}, \bibinfo {author}
  {\bibfnamefont {H.~C.}\ \bibnamefont {Kapteyn}}, \ and\ \bibinfo {author}
  {\bibfnamefont {M.~M.}\ \bibnamefont {Murnane}},\ }\bibfield  {title}
  {\enquote {\bibinfo {title} {Temporal phase control of soft-x-ray harmonic
  emission},}\ }\href@noop {} {\bibfield  {journal} {\bibinfo  {journal} {Phys.
  Rev. A}\ }\textbf {\bibinfo {volume} {58}},\ \bibinfo {pages} {R30} (\bibinfo
  {year} {1998})}\BibitemShut {NoStop}%
\end{thebibliography}

%

\end{document}